\documentstyle[amssymb,aps,multicol,prl,epsf]{revtex}
\newcommand \beq  {\begin{equation}}
\newcommand \eeq  {\end{equation}}
\newcommand \bea {\begin{eqnarray} }
\newcommand \eea {\end{eqnarray}}

\begin{document}
%\draft
\title{Superconductivity, phase fluctuations and the c-axis conductivity of bilayer high temperature
superconductors}
\author{N. Shah and A. J. Millis}
\address{Center for Materials Theory\\
Department of Physics \& Astronomy, Rutgers University\\
136 Frelinghuysen Road, Piscataway, NJ 08854}
%\date{\today}
\maketitle

\begin{abstract}
We present a theory of the interplane conductivity of bilayer high temperature superconductors, 
focusing on the effect of quantal and thermal fluctuations on the oscillator strengths 
of the superfluid stiffness and the bilayer plasmon. We find that the opening of the superconducting
gap and establishment of superconducting phase coherence each lead to redistribution
of spectral weight over wide energy scales. The factor-of-two relation between
the superfluid stiffness and the change below $T_c$ in the oscillator strength
of the absorptive part of the conductivity previously derived for single-layer
systems, is found to be substantially modified in bilayer systems.
\end{abstract}

\begin{multicols}{2}

\section{Introduction}

The interlayer (`c-axis') conductivity of high temperature superconductors
is an important and long standing problem. Experimental 
results \cite{Ong91,Homes93,Marel98,Basov98,Basov00} have seemed to many workers
\cite{Anderson93,Anderson96,Clarke96,Chakravarty96} to be sharply at variance
with conventional understanding and to imply the existence of radically new
physics. Other workers, conversely, have argued that many aspects of the
results can be understood in a straightforward manner \cite{Hwang98}. 
Especially interesting have been apparent violations of the 
Ferrel-Glover-Tinkham sum rule\cite{Tinkham75} relating the superfluid stiffness 
to changes in the absorptive part of the conductivity as temperature is 
reduced below the transition temperature $T_{c}$.

Recently Ioffe and one of us \cite{Ioffe99,Ioffe00} have argued that the
interlayer conductivity is a theoretically simple object (basically the
convolution of two in-plane Green functions) and is therefore a sensitive
probe of in-plane scattering rates and of the quantal and thermal phase
fluctuations characterizing the superconducting state. A number of predictions 
were made, some of which appear to
agree with experiment and some of which do not\cite{Basov00}. 
The results reported in \cite{Ioffe99,Ioffe00} had a crucial limitation: 
the equations were derived for a `single-
layer' system such as $La_{2-x}Sr_{x}CuO_{4}$, whereas most (but not all) of
the experimentally studied systems (including $Bi_{2}Sr_{2}Ca_{2}Cu_{2}O_{8}$
and $YBa_{2}Cu_{3}O_{7-\delta }$) have a `bilayer' structure, with a unit cell
containing two superconducting $CuO_{2}$ planes coupled to each other more
strongly than to the planes in adjoining unit cells. The new feature introduced 
by the bilayer structure is ``local field corrections'': 
application of a uniform field can lead to a non-uniform 
charge distribution within a unit cell, which in turn 
causes internal fields affecting the motion of charges. 
This leads to phenomena not found in single plane systems; for example, the
bilayer plasmon feature observed and discussed by van der Marel and others \cite{Marel98}%
. Interest in this feature was recently increased by the observation \cite{Marel00} 
that the bilayer plasmon frequency  may provide information about the 
in-plane electronic compressibility, a quantity of great theoretical 
interest not easily accessible by other techniques.

In this paper we generalize the treatment of \cite{Ioffe99,Ioffe00} to the
bilayer case. We provide a simple and physically transparent treatment of 
the c-axis conductivity in the limit (appropriate for the high
temperature superconductors) in which the interplane coupling is weak 
relative to in-plane energy scales. Our treatment includes phonon, bilayer
plasmon and quasiparticle absorption. Our results 
provide a justification for previously proposed phenomenological oscillator
models and allow us to determine the interplay between bilayer plasmon features and
interlayer phase coherence. Our methods may easily be generalized to more
complicated situations such as the three and four layer structures found in
other high-$T_{c}$ materials, but this generalization is not given here. 

The rest of this paper is organized as follows. 
In Sec II we present the formalism; Sec III gives results calculated in the absence
of phonons; Sec IV discusses the spectral weight and sum-rule analysis. 
In Sec V we extend our treatment to incorporate phonons (relevant for some bilayer materials) 
and finally in Sec VI we summarize our conclusions and discuss the applications to experiment.
\section{Formalism}

\subsection{Fundamental Equations}

We study the bilayer system shown in Fig 1 in which each unit cell contains
two conducting planes separated by a distance $d_{1}$ and coupled by a
hopping $t_{1}$. The distance between a plane in one unit cell and the
closest conducting plane in another unit cell is $d_{2}$, so the lattice
parameter in the interplane direction is $d=d_{1}+d_{2}$. Planes separated by a distance 
$d_{2}$ are coupled by a hopping $t_{2}$. We neglect further neighbor hoppings, although 
these  can be easily
added at the cost of increased complexity of our equations. In the high $
T_{c}$ context, $t_{2}<<t_{1}$ (but our results are valid for any ratio $t_{2}/t_{1}$) 
and both  $t_{1}$ and $t_{2}$ depend strongly on in-plane momentum, 
being maximal for momenta in the ($0,\pi $) region of the zone and 
minimal for momenta near the zone diagonals $(\pm \pi,\pm \pi)$. 
We will usually not write the momentum dependence explicitly.

%system3.eps

\vspace{0.25cm} 
\centerline{\epsfxsize=3truein \epsfbox{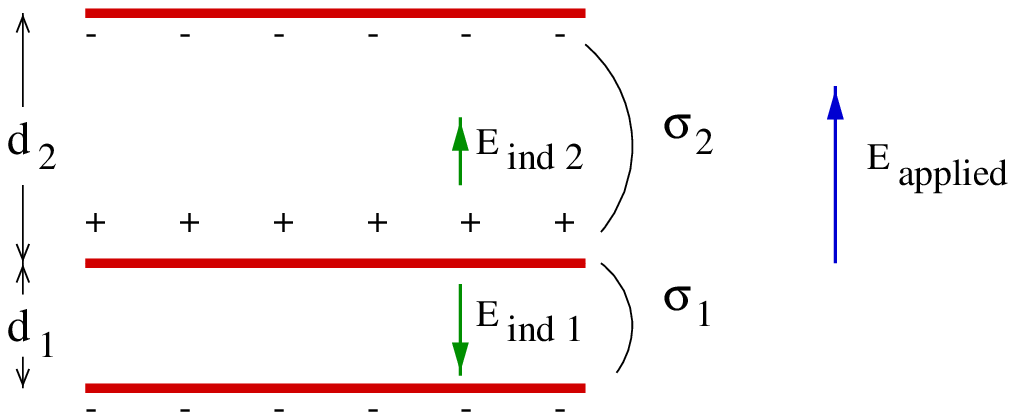}}
{\footnotesize \textbf{Fig. 1} Geometry considered in present paper. Shown is 
one unit cell (chosen as two close planes) and part of the next unit cell. 
The figure also shows the applied spatially uniform electric field ($E_{applied}$), 
the charge build-up on planes (represented as $++++$ and $----$
and the resultant induced fields $E_{ind 1,2}$ which in turn affect the charge flow.}
\vspace{0.25cm}

We refer to the two planes in one unit cell by the index $l=a,b$ and to the region between 
two planes separated by $d_{1}$ as region $1$ and the region between two planes separated by $d_{2}$
as region $2$; we label the unit cells by the index $i$. \ We take all
planes to be identical and neglect all interplane couplings except for the
hoppings and the internal electric fields induced by nonuniform charge
distributions. We allow for the possibility that the planes are at different
electrochemical potential $\mu $. The Hamiltonian describing the system is
then 
\begin{eqnarray}
&&H=\sum_{i,l}H_{in-plane}\\
&&-\sum_{i,\sigma }\int \frac{d^{2}p}{(2\pi )^{2}}%
t_{1}(p)\left( e^{i\left( \mu _{i,a}-\mu _{i,b}\right) t}c_{i,a,p,\sigma
}^{+}c_{i,b,p,\sigma }
+H.c.\right) -
\nonumber \\
&&\sum_{i,\sigma }\int \frac{d^{2}p}{(2\pi )^{2}}t_{2}(p)\left( e^{i\left(
\mu _{i-1,b}-\mu _{i,a}\right) t}c_{i-1,b,p,\sigma }^{+}c_{i,a,p,\sigma
}
+H.c.\right),
\nonumber
\end{eqnarray}
where $H_{in-plane}$(which we will not need to specify) describes the electronic physics within a  $CuO_{2}$ plane. 
We shall study the properties of this Hamiltonian by a perturbation
expansion in $t_{1}$ and $t_{2}$. The dimensionless parameter is $
t_{1,2}/ E_{in-plane}$ where $E_{in-plane}$ is the in-plane density of states or inverse 
of some other important in-plane local energy scale. This
approach has been shown to agree with results obtained by other means in a
number of contexts, including coupled Luttinger liquids \cite
{Georges00,Schofield01} and semiconductor heterostructures \cite
{Millis01}. 

We are interested in optical experiments \cite
{Homes93,Marel98,Basov98,Basov00} which may be thought of as involving the
application to the system of a weak spatially uniform transverse electric
field of magnitude $E_{T}$ directed perpendicular to the planes. The
experimentally determined quantity is the bilayer conductivity $\sigma
_{bilayer}$, which is the coefficient relating the applied electric field to
the spatial average of the current. The applied field leads to an
electrochemical potential $\mu _{i,l}$ on each plane which has three
contributions: from the applied electric field, from fields generated by
build-up of charge on particular planes (shown as $E_{ind 1,2}$ in Fig 1), 
and from changes in the in-plane
chemical potential due to changes in the in-plane density. We have 
\begin{eqnarray}
\mu _{i,l}=eE_{T}R_{i,l}+eV_{ind}[\{n_{i^{\prime },l^{\prime }}\}]+\chi^{-1}
n_{i,l}/e^{2},
\end{eqnarray}
where $R_{i,l}$ is the position vector of the plane in the interplane 
direction and the zeros of charge density $n$ and of chemical
potential $\mu$ have been defined to correspond to the states of the planes in equilibrium. $
eV_{ind} $ is the electrochemical potential due to electric fields produced
by charge build up and $\chi^{-1} =e\partial \mu /\partial n$ is inverse of the the exact in-plane 
density-density correlation function of $H_{in-plane}$. The factors of $e$
arise from converting particle densities to charge densities.

The spatially varying chemical potential leads to interplane electrical
currents described by operators such as 
\begin{equation}
j_{i,1}=-et_{1}\sum_{\sigma }\int \frac{d^{2}p}{(2\pi )^{2}}\left(
e^{ieVi,_{1}t}c_{i,ap\sigma }^{+}c_{i,bp\sigma }-H.c.\right)
\end{equation}
and therefore to interplane charge build-up, for which we must solve self
consistently. \ In the present simple situation, application of a uniform
electric field leads to two independent densities $n_{a}$ and $n_{b}$ and
two independent chemical potential differences, $eV_{1}=\mu _{i,a}-\mu
_{i,b} $ and $eV_{2}=\mu _{i-1,b}-\mu _{i,a}$. Combining the continuity
equation for the current, the Maxwell equation and the density, and
evaluating the currents to leading nontrivial order in $t_{1,2}$ and $E_{T}
$ leads to an expression for $\sigma _{bilayer}$. This expression is most
conveniently written in terms of conductivities $\sigma _{1}$ ($\sigma _{2}$%
) appropriate to a `single-layer' material consisting of an infinite stack
of identical planes all separated by distance $d_{1}$ ($d_{2}$) and coupled
by hopping $t_{1}$ ($t_{2}$) and is 
\begin{equation}
\sigma_{bilayer}(\omega )=\frac{\sigma _{1}\sigma _{2}-i\omega (\sigma
_{1}\widetilde{d}%
_{1} +\sigma _{2}\widetilde{d}_{2} )/C}{\sigma _{1}\widetilde{d}_{2}+\sigma _{2}\widetilde{d}%
_{1}-i\omega /C}  
\label{sigmabilayer}
\end{equation}
with $\widetilde{d}_{1,2}=d_{1,2}/(d_{1} + d_{2})$. Here the ``blockade parameter''
\begin{equation}
C=\frac{4\pi }{\varepsilon }+\frac{2\chi d}{e^{2}d_{1}d_{2}},
\label{C}
\end{equation}
where
$\varepsilon $ is the
`background' dielectric function due to non-electronic degrees of freedom. 
In this paper we shall take $\varepsilon $ to be constant except in Sec V where we note include 
the effects of phonons (important in the optical absorption of some high-$%
T_{c}$ materials) by using an $\varepsilon (\omega )$ with
appropriate frequency dependence in the expressions for $\sigma _{bilayer}$.  
The constant $C$ expresses the blocking effects arising 
because charge which flows onto a plane via the strong link (large
conductivity) will take a long time to flow off via the weak link: if the
driving frequency is low, charge buildup will therefore occur, inhibiting
additional motion of charge across the strong link.

Eq \ref{sigmabilayer} reproduces all of the obvious limits correctly: as $\omega
\rightarrow 0$ it reduces to 
$\sigma_{bilayer}^{-1} = \widetilde{d}_{2}/ \sigma_{2} + 
\widetilde{d}_{1}/ \sigma_{1}$ so $\sigma _{bilayer}$ is dominated by 
smaller of the conductivities as expected, while if $\omega = C%
%TCIMACRO{\func{Im}}%
%BeginExpansion
\mathop{\rm Im}%
%EndExpansion
\left( \sigma _{1}(\omega)\widetilde{d}_{2}+\sigma _{2}(\omega)\widetilde{d}_{1}\right) $
then a `bilayer plasmon' pole occurs (damped, of course, by the dissipative
part of the conductivity). If $d_{1}=d_{2}$ and $\sigma _{1}=\sigma
_{2}$ then the system becomes effectively single-layered and Eq  \ref{sigmabilayer} 
shows $\sigma_{bilayer} = \sigma_1$.

The calculation of the constituent conductivities $\sigma _{1,2}$ is given
in \cite{Ioffe99,Ioffe00} and relevant results will be recalled in the next
subsection. We note here that for consistency they (and $\chi^{-1} $) must be
calculated to leading nontrivial order in the interplane hoppings $t_{1,2}$. $\chi^{-1} $ is therefore a single-plane quantity and $\sigma_{1,2}$ may therefore be expressed in terms of
convolutions of two-dimensional in-plane Green functions. If higher order expressions are used then for example, exchange
interaction contributions must be included in $C$ and further changes to $%
\sigma _{bilayer}$ will occur.

We see that the frequency dependence of the bilayer conductivity is
complicated and depends on the value of $C$ and on the magnitudes and
frequency dependences of the individual conductivities. In 
general $\sigma _{bilayer}$ exhibits three regimes: a high frequency
regime in which $\sigma _{bilayer}=\sigma _{1} \widetilde{d}_{1} +\sigma _{2}\widetilde{d}_{2}$; a low frequency
regime in which $\sigma _{bilayer}\approx \sigma _{2}/\widetilde{d}_{2}$ and
a broad crossover regime with characteristic scale 
\begin{equation}
\omega^*=C \left | \sigma _{1}(\omega^*)\widetilde{d}_{2}+\sigma _{2}(\omega^*)%
\widetilde{d}_{1} \right |  
\label{w*}
\end{equation} 
which depends on the conductivities. If in the superconducting state, $\omega^* < 2\Delta$ 
then the scale $\omega^*$ becomes identical to the bilayer plasmon frequency 
$\omega_{bilayer}$ and near $\omega_{bilayer}$ we have 
\begin{equation}
\sigma(\omega \sim \omega_{bilayer} < 2\Delta) = \frac{-i \rho_{bilayer}}
{\omega - \omega_{bilayer} + i \delta}
\label{rhob}
\end{equation}
defining the strength $\pi\rho_{bilayer}$ of the bilayer plasmon absorption.

\subsection{Constituent conductivities}

The calculation of the constituent conductivities is discussed at length in 
\cite{Ioffe99,Ioffe00}. Here we briefly recall key results and needed
formulae. The conductivities are given by correlation functions
of current operators such as $j_{1}$ above and involve expectation values of
the form $t^{2}(p)<c_{i,p}^{+}(t)c_{j,p}(t)c_{j,p}^{+}(t^{\prime
})c_{i,p}(t^{\prime })>$ $.$ To leading order in $t$, correlations between
operators on different planes vanish so the expression may be written as the
sum of two terms, one involving $<c_{i,p}^{+}(t)c_{i,p}(t^{\prime
})><c_{j,p}^{+}(t^{\prime })c_{j,p}(t)>$ (i.e. the product of two `normal'
in-plane Green functions $G(p,t-t^{\prime })$) and one involving $%
<c_{i,p}^{+}(t)c_{i,p}^{+}(t^{\prime })><c_{j,p}(t^{\prime })c_{j,p}(t)>$
(i.e. the product of two `anomalous' in-plane Green functions $%
F(p,t-t^{\prime })$). However, the anomalous Green function involves the
superconducting order parameter which has a phase which we denote by $\phi $
. The product of anomalous Green functions on planes $i$ and $j$ therefore
involves the factor $e^{i\left( \phi _{i}(r,t\right) -\phi _{j}(r^{\prime
},t^{\prime }))}$  (times a short ranged function of $r,t$ which depends on
the details of the interplane hopping and the underlying energy scales of
the superconductivity) and must be averaged over an ensemble describing the
phase fluctuations. Refs \cite{Ioffe99,Ioffe00} showed that in in the case
of interest here, these effects may be accounted for by multiplying the $%
F-F^{+}$ contribution to $\sigma $ by a Debye-Waller factor $\alpha $ which
is unity for a mean-field BCS superconductor with no fluctuations, may be
reduced from unity by quantal or thermal fluctuations about an ordered
state, and which becomes very small if there is no long range phase
order. We follow refs \cite{Ioffe99,Ioffe00} in assuming that the pseudogap 
state is characterized by a conventional superconducting 
gap but no interplane phase coherence. 

Thus ($\nu =1,2$ labels planes)
\beq
\sigma _{\nu }(i\omega _{n})=\frac{K_{\nu }+ \Pi_{\nu}}{i\omega } 
\label{sigmanu}
\eeq
with the diamagnetic contribution  given by 
\bea
K_{\nu }&=&4e^{2}d_{\nu }T\sum_{n}\int \frac{d^{2}p}{\left( 2\pi \right) ^{2}}
t_{\nu }(p)^{2}
\nonumber \\
&&\left(-G(p,\omega^{'}_{n})G(p,\omega^{'} _{n})
+\alpha F(p,\omega^{'}
_{n})F(p,\omega^{'} _{n})\right)
\eea
and the paramagnetic contribution given by
\bea
&&\Pi_{\nu}=4e^{2}d_{\nu }T\sum_{n}\int \frac{d^{2}p}{\left( 2\pi \right) ^{2}}t_{\nu }(p)^{2}
\nonumber \\
&&\left(
G(p,\omega _{n}+\omega^{'} _{n})G(p,\omega^{'} _{n})
+\alpha F(p,\omega _{n}+\omega^{'}
_{n})F(p,\omega^{'} _{n})\right). 
\eea
where $G,F$ are the exact normal and anomalous Green
functions corresponding to $H_{in-plane}$).

The $\omega \rightarrow 0$ limit is $\sigma \rightarrow \frac{i\rho _{s}}{%
\omega }$ with 
\begin{equation}
\rho _{s,\nu }=8\alpha e^{2}d_{\nu }T\sum_{n}\int \frac{d^{2}p}{\left( 2\pi
\right) ^{2}}t_{\nu }(p)^{2}F(p,\omega^{'} _{n})F(p,\omega^{'} _{n})
\end{equation}
while the usual `f-sum rule' arguments \cite{Kohn63} yield ($\sigma ^{(1)}$ is
the absorptive part of the conductivity) 
\begin{equation}
K=\int_{0}^{\infty }\frac{2d\omega }{\pi }\sigma ^{(1)}(\omega )=\rho
_{s}+\int_{0^{+}}^{\infty }\frac{2d\omega }{\pi }\sigma ^{(1)}(\omega )
\end{equation}
(note that in the first equality the integral is only over one half of the delta
function at $\omega =0$).

In the high-$T_{c}$ materials the anisotropy of $t(p)$ is such that the interplane 
conductivity is dominated by the ``corner'' regions around $(0,\pi
)$ and so we follow refs \cite{Ioffe99,Ioffe00} and neglect both the angular
variation of the gap and of $t_{1,2}.$ In this approximation $\sigma _{1}$
and $\sigma _{2}$ have the same frequency dependence and differ only by a
prefactor involving the square of the relevant hopping.

In high-$T_{c}$ materials, the normal state c-axis conductivity is
characterized by a very broad Drude-like absorption, corresponding to an
in-plane Green function (in the ``corner region'') characterized by a very
large, essentially frequency independent scattering rate. We therefore follow 
refs \cite{Ioffe99,Ioffe00} and use
this `dirty-limit' form to compute the conductivities. We note that there is
a large and growing literature on changes to the Green function as the
temperature is changed through the superconducting transition \cite
{Norman98,Chubukov99,Shen99}. The implication of these changes for the
c-axis conductivity has been studied by \cite{Ioffe00}, but because our main
interest here is in the new features introduced by the bilayer structure
and because there is no consensus on the physical origin or mathematical
form of the superconductivity-induced changes, we do not consider them here. 
Further, we shall be interested
mainly in three situations--the normal state, at a temperature well above
the `pseudogap formation temperature', the $T\rightarrow 0$ limit in the
superconducting state, and temperatures well below the pseudogap scale and
near to $T_c$, i.e. $T_c<T<<\Delta$. Thus we may neglect the temperature,
except as it influences the value of the Debye-Waller parameter $\alpha $. We therefore have
\begin{eqnarray}
\sigma _{\nu }^{(1)}(\omega )&=&-\sigma _{0,\nu }\Theta (\left| \omega
\right| -2\Delta )\\
&&\int_{\Delta }^{\left| \omega \right| -\Delta }\frac{%
\omega^{'} (\omega^{'} -\left| \omega \right| )+\alpha _{\nu }\Delta ^{2}}{\sqrt{{%
\omega^{'} }^{2}-\Delta ^{2}}\sqrt{{(\omega^{'} -}\left| {\omega }\right| {)}%
^{2}-\Delta ^{2}}}\frac{d\omega^{'} }{\left| \omega \right|}
\nonumber \\
\sigma _{\nu }^{(2)}(\omega )&=&\frac{\Delta (\alpha _{\nu}-1)\pi }{2\left| \omega\right|} 
+ \sigma _{0,\nu }sgn(\omega )\Theta (2\Delta
-\left| \omega \right| ) 
\\
&&\int_{\Delta }^{\left| \omega \right| +\Delta
}\frac{\omega^{'} (\omega^{'} -\left| \omega \right| )+\alpha _{\nu }\Delta ^{2}}{%
\sqrt{{\omega^{'} }^{2}-\Delta ^{2}}\sqrt{\Delta ^{2}-{(\omega^{'} -}\left| {\omega }%
\right| {)}^{2}}}\frac{d\omega^{'} }{\left| \omega \right| } 
\nonumber \\
&&+\sigma _{0,\nu }sgn(\omega )\Theta (\left| \omega \right| -2\Delta
)
\nonumber \\
&&\int_{\left| \omega \right| -\Delta }^{\left| \omega \right| +\Delta
}\frac{\omega^{'} (\omega^{'} -\left| \omega \right| )+\alpha _{\nu }\Delta ^{2}}{%
\sqrt{{\omega^{'} }^{2}-\Delta ^{2}}\sqrt{\Delta ^{2}-{(\omega^{'} -}\left| {\omega }%
\right| {)}^{2}}}\frac{d\omega^{'} }{\left| \omega\right| } \nonumber  
\end{eqnarray}
with $\sigma _{0,\nu }$ the normal state (neither superconductivity nor any
gap) conductivity, frequency independent because we have taken the dirty
limit. We note that because $t_{1,2}$ differ, so also may the quantal
fluctuation parameters $\alpha _{1,2}$. The considerations of \cite{Ioffe00}
suggest that $\alpha$ is dominated by in-plane fluctuations, so may not differ
much between the two links, so in the rest of this paper we set
$\alpha_1=\alpha_2$.

The single-layer superfluid stiffnesses following from these expressions are 
\begin{equation}
\rho _{s,\nu }= \alpha _{\nu } \rho _{s,\nu }^{BCS} = \alpha _{\nu } \pi\sigma _{0,\nu }\Delta,
\label{rhoBCS}
\end{equation}
where  $\rho _{s,\nu }^{BCS}$ is the superfluid stiffness following from the assumption 
of full phase coherence$(\alpha_{\nu}=1)$.

\section{Calculated conductivity}

In this section we evaluate the formulas derived in the previous section.
The fundamental result was Eq \ref{sigmabilayer}, which
expressed the conductivity $\sigma _{bilayer}$ of a bilayer system in terms
of the conductivities $\sigma _{1,2}$ of effective `single-layer' systems
corresponding to the two interplane spacings of the bilayer and a
``blockade parameter'' $C$ expressing interplane interaction effects. 
We use the normal state ($\Delta=0$) value of the characteristic frequency 
scale $\omega^{*}$ defined by Eq \ref{w*}, 
\begin{equation}
\omega^{*}= C\sigma _{0,1}\widetilde{d}_{2}\left( 1+\frac{\sigma _{0,2}%
\widetilde{d}_{1}}{\sigma _{0,1}\widetilde{d}_{2}}\right),  \label{omegabilayernorm}
\end{equation}
in our discussion henceforth. It is most convenient to
express $C$ in terms of this normal state value of $\omega^{*}$. 

The important dimensionless parameters are the ratio of normal state conductivities 
$b=\sigma _{0,2}/\sigma _{0,1}<1 $%
, $\omega^{*}/\Delta $ \ and the Debye-Waller
factors $\alpha _{1,2}$ introduced above Eq \ref{sigmanu}. To simplify the
presentation of our results, we define conductivity units such that $\sigma
_{0,1}=1$ and frequency units such that $\Delta = 1$. For definiteness we 
choose the normalized interplane distances to be $\widetilde{d}_{1}=0.4$, 
$\widetilde{d}_{2}=0.6$ so $\omega^{*}= 0.6 \sigma _{0,1} C(1+ 2b/3).$ and set $\alpha_1=\alpha_2$.

\vspace{0.25cm} 
\centerline{\epsfxsize=3.2truein \epsfbox{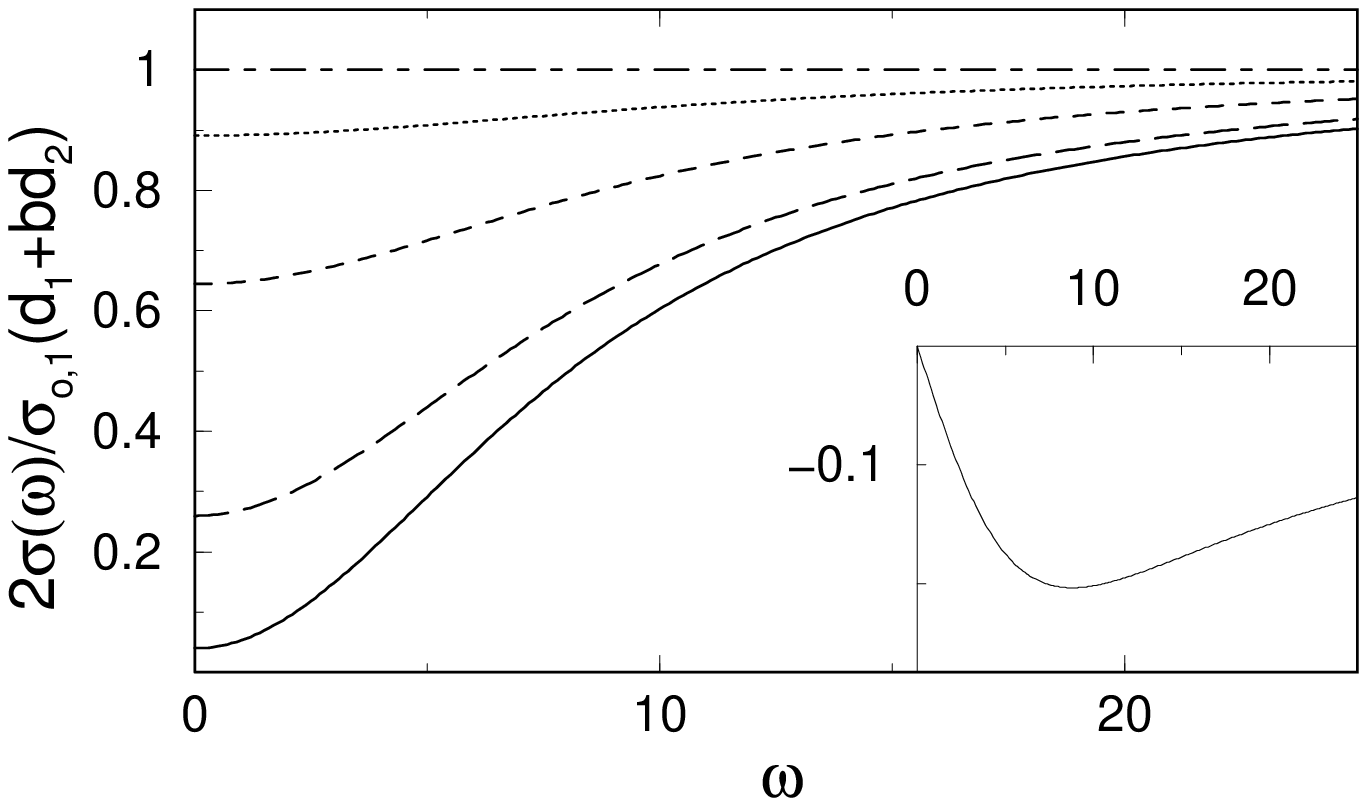}}
{\footnotesize \textbf{Fig. 2} Real part of normal state conductivity scaled by $\sigma _{0,1}(\widetilde{d}_{1} + b\widetilde{d}_{2}) $ for 
$b=1,0.5,0.25,0.075,0.01 $ (from top to bottom). The bilayer frequency 
$\omega^* = 11.48, 9.91, 8.8, 8.39$ for $b=0.5,0.25,0.075,0.01$, respectively. 
The inset shows imaginary part of the conductivity for $b=0.075$.
 }
\vspace{0.25cm}

Fig 2 shows the real part of the normal state conductivity for the five
values $b=1,0.5,0.25,0.075,0.01 $. The suppression of the low frequency conductivity by
the blockade effect is evident, as is the gradual crossover to the high
frequency isolated layers value. The curves have been scaled by $\sigma _{0,1}(\widetilde{d}_{1} + b\widetilde{d}_{2})$ so that all have the same high frequency limit. The inset shows the imaginary part for $b=0.075$. In the
crossover regime, the blockade effect is seen to lead to out of phase
response.

We now consider the superconductivity induced changes. The top panel of
Fig 3 shows the result of evaluating 
Eq \ref{sigmabilayer} in the `single-layer' ($b=1$) fully phase coherent $(\alpha_{\nu}=1)$ case. 
The opening of the superconducting gap suppresses the real part of the
conductivity for frequencies below $2\Delta $ and changes the form somewhat
for $\omega \gtrapprox 2\Delta $. The establishing of superconducting 
phase coherence leads to a divergent low frequency response characterized 
by the superfluid stiffness $\rho_{s,bilayer}$. The oscillator strength in
the superfluid response is shown in the top panel as a shaded rectangle. 
The f-sum rule arguments discussed at length in the next section imply that in the 
fully phase coherent $(\alpha_{\nu}=1)$ case, the area lost in the absorptive part of $\sigma$
due to the opening of the superconducting gap is transferred to the superfluid response.
It is apparent from the figure that the area in the shaded rectangle is approximately
equal to the 'missing' area and we have verified numerically that the areas are equal:
$\int_{0}^{\infty} d\omega [\sigma(\Delta = 0)-\sigma(\Delta)] = \pi \rho _{s,bilayer}/2$.
%was fg3.eps

\vspace{0.25cm} 
\centerline{\epsfxsize=3.2truein \epsfbox{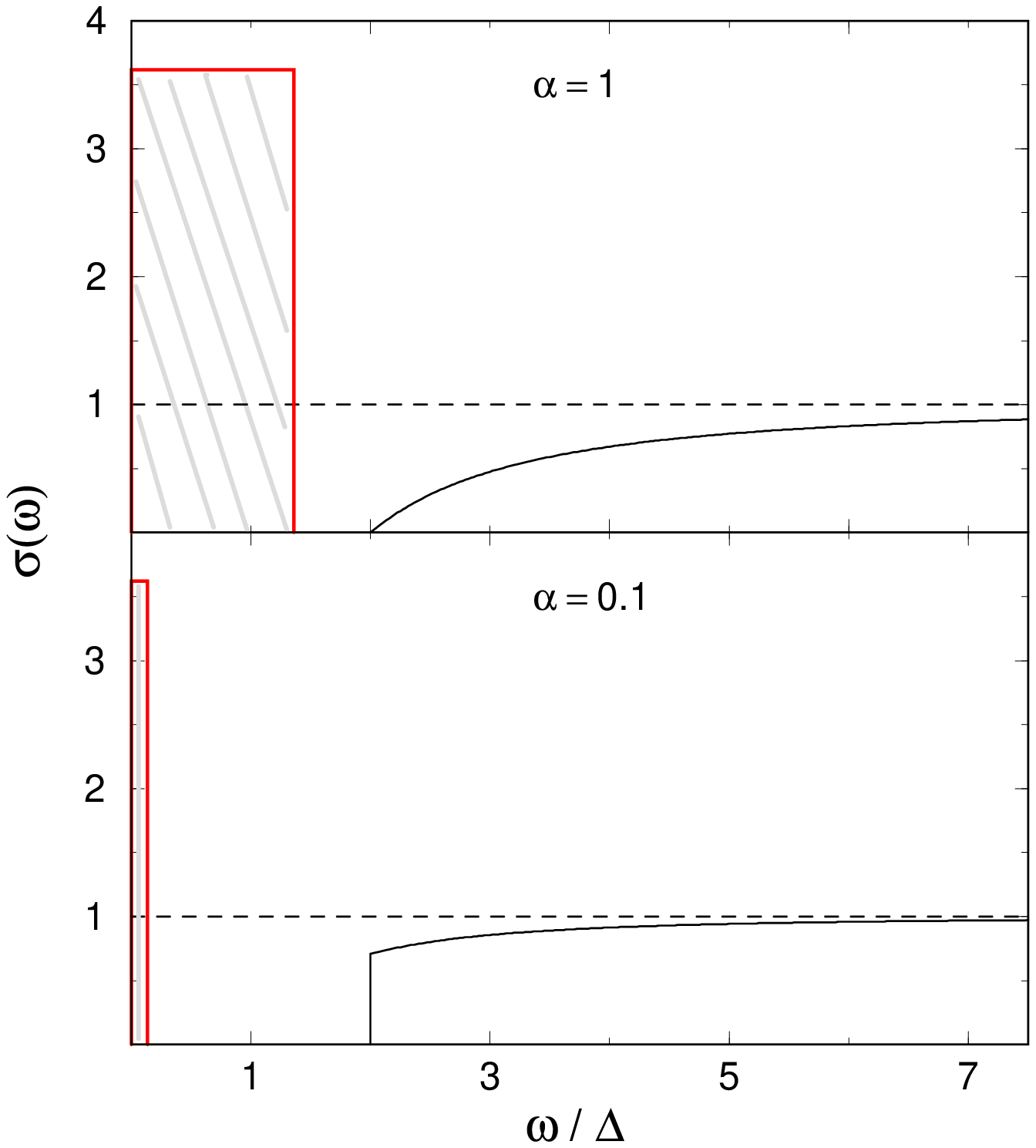}}
{\footnotesize \textbf{Fig. 3} Real part of conductivity for `single-layer'$(b=1)$ system. Top panel corresponds to fully phase coherent ($\alpha_\nu = 1$) while the bottom panel to $\alpha_\nu = 0.1$ superconducting state. The dashed line in both corresponds to the normal state. The area of the shaded rectangle equals $\pi \rho_{s} /2$ in each case. }
\vspace{0.25cm}

If $b<1$ then the situation is much more involved. In particular because the form
of the conductivity in the regime $\omega \sim \omega^{*}$ depends
sensitively on the interplay between $C$ and $\sigma _{1,2}$, the superconductivity 
induced changes in $%
\sigma _{1,2}$ will lead to large changes in $\sigma _{bilayer}$. In addition,
some of the oscillator strength eliminated from the $\omega<2\Delta$ region by
the opening of the gap  will go into the bilayer plasmon absorption instead
of into the superfluid delta function.

\vspace{0.25cm} 
\centerline{\epsfxsize=3.2truein \epsfbox{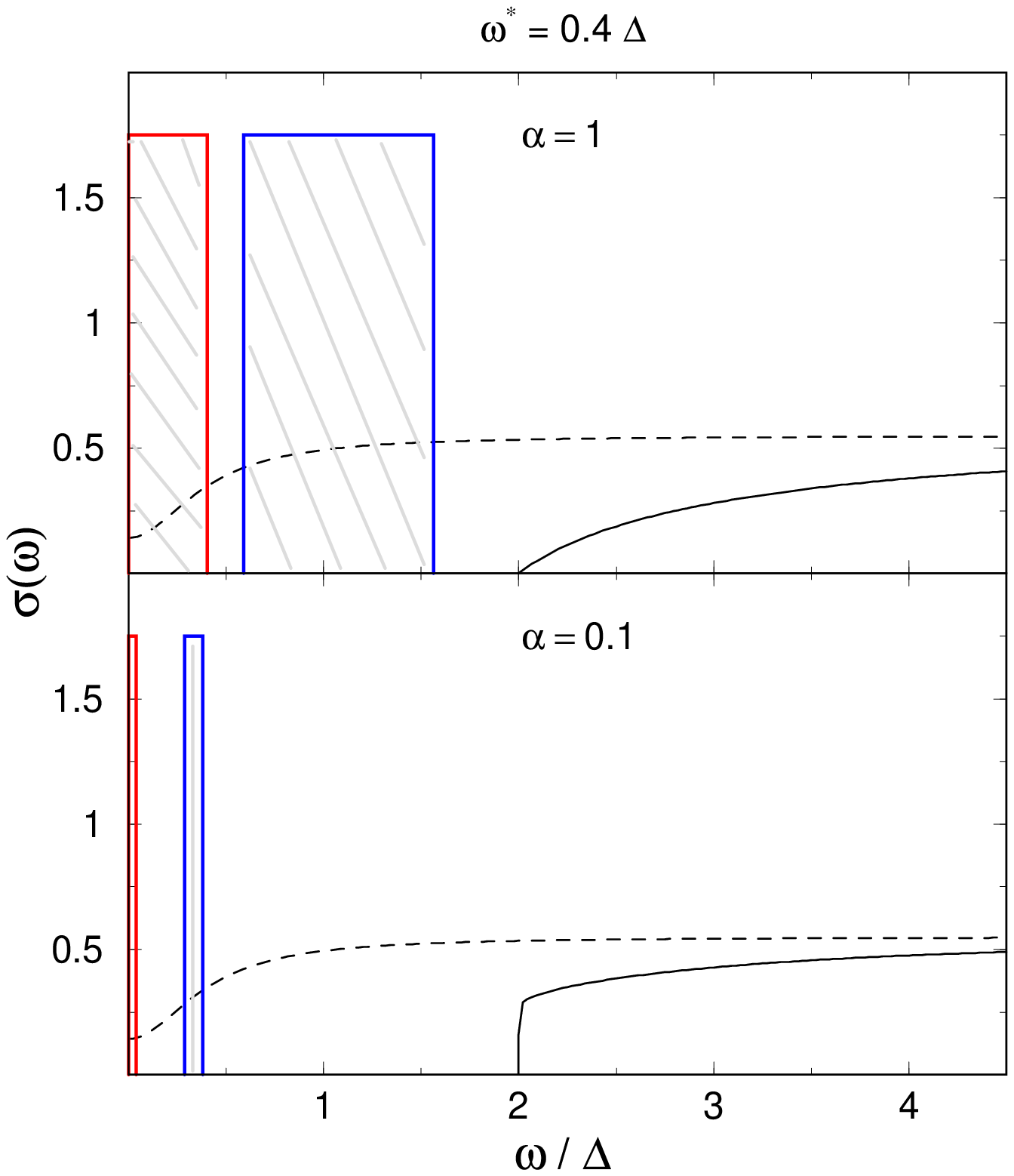}}
{\footnotesize \textbf{Fig. 4} Real part of conductivity for bilayer 
$(b=0.075)$ system for $\omega^* =0.4 \Delta$. Top panel corresponds 
to fully phase coherent ($\alpha_\nu = 1$) and  bottom panel to 
$\alpha_{\nu}=0.1$  superconducting state. The dashed line in both 
corresponds to the normal state conductivity. The area of the two 
shaded rectangles in each panel equals $\pi \rho_s /2$ and 
$\pi \rho_{bilayer}$ for rectangles centered at $\omega=0$ and  
$\omega_{bilayer}$, respectively. }
\vspace{0.25cm}

The top panels of Figs 4, 5 and 6 show the normal and superconducting state conductivity in the
fully phase coherent ($\alpha_{\nu}=1$) limit for three representative values of $\omega^{*} 
$: $0.4\Delta $, $3\Delta$ and $10\Delta $, respectively  in the strongly
anisotropic limit $b=0.075$. The oscillator strengths in the superfluid response
and (if it is inside the gap) the bilayer plasmon are shown as shaded rectangles. For 
$\omega^{*}=0.4\Delta $, the bilayer plasmon lies within the superconducting gap. 
The resulting absorption is a delta function at the marked frequency,
with an intensity corresponding to an integrated area equal to that of the
rectangular box shown. For $\omega^{*}= 3\Delta$ the bilayer plasmon 
lies just above the gap, visible as a sharp feature at the gap edge and 
the remainder of the superconducting conductivity is slightly suppressed over 
a wide frequency range. For $\omega^{*}= 10 \Delta$ the bilayer plasmon
feature is evident only as a very broad absorption at frequencies that 
are much larger than shown here. 

\vspace{0.25cm} 
\centerline{\epsfxsize=3.2truein \epsfbox{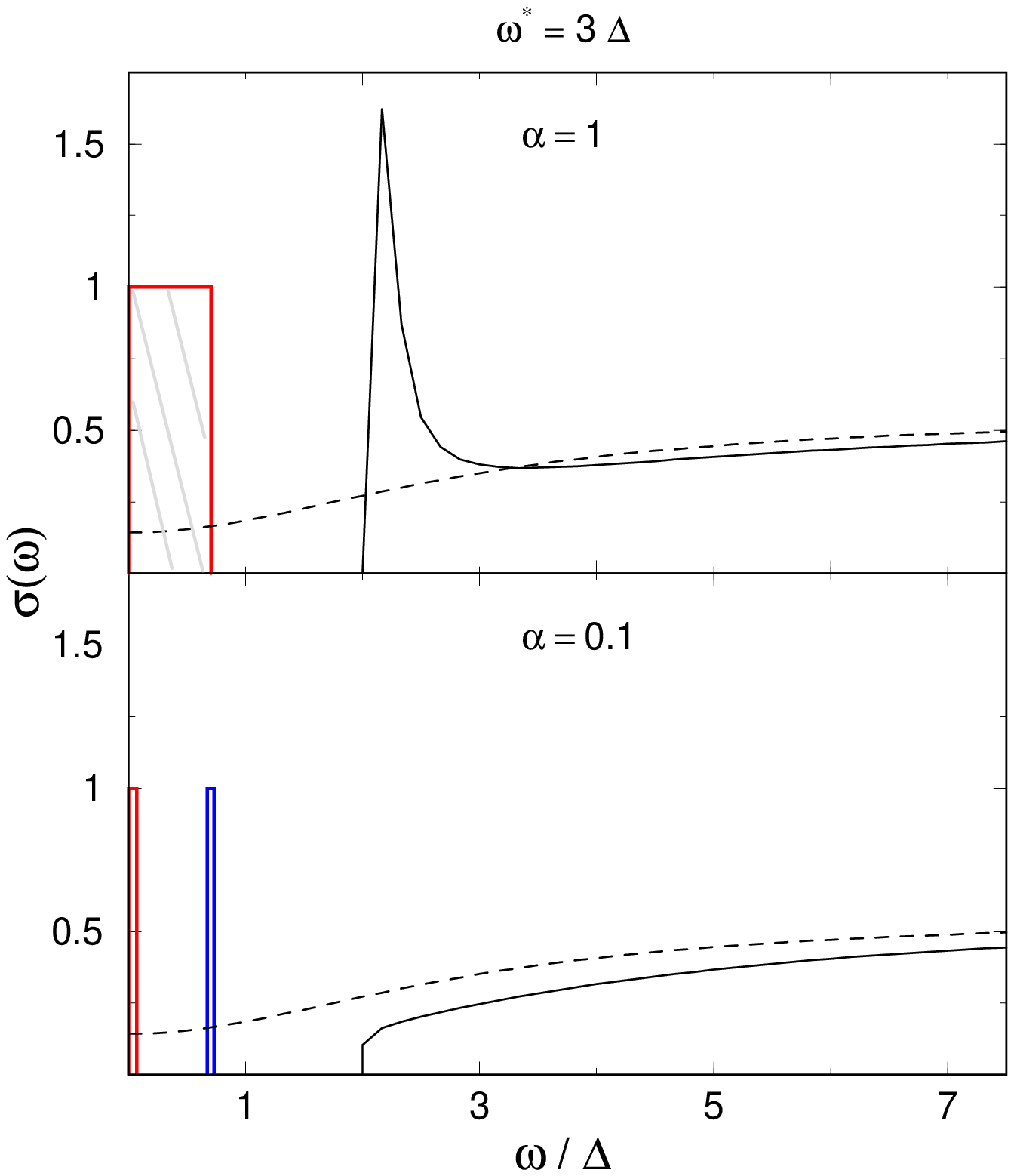}}
{\footnotesize \textbf{Fig. 5} Real part of conductivity for 
bilayer $(b=0.075)$ system for $\omega^* = 3 \Delta$. Top panel 
corresponds to fully phase coherent ($\alpha_\nu = 1$) and  bottom 
panel to $\alpha_{\nu}=0.1$  superconducting state. The dashed line 
in both  corresponds to the normal state conductivity. The area of the 
shaded rectangle at $\omega=0$ in each panel equals $\pi \rho_s /s$ while 
the area of the rectangle at $\omega_{bilayer}$ in the second panel  equals 
$\pi \rho_{bilayer}$.}
\vspace{0.25cm}

The bottom panels of Figs 4, 5 and 6 show the effect of phase fluctuations, displaying the
superconducting curves for the same $\omega ^{*}$ and $b$ values as the respective top
panels, but with $\alpha_{\nu}=0.1$. For $\omega > 2\Delta$ the difference 
between the normal and the superconducting conductivity increases as 
phase fluctuations become more important. Both the
strength and the frequency of the bilayer plasmon feature depend strongly on
the value of the fluctuation parameter. For $\omega^{*} =  3\Delta,10\Delta $, 
the bilayer plasmon moves below the gap as seen in Figs 5 and 6. For the 
pseudogap case ($\alpha_\nu = 0$) both the bilayer plasmon and the superconducting 
delta function are absent and the $\omega > 2\Delta$ conductivity is roughly 
the same as the $\alpha_\nu = 0.1$ case. We can also study the conductivity 
for different values of $\alpha_1$ and $\alpha_2$ and it is worth
noting that in the case when $\alpha_2=0$ and $\alpha_1 \ne 0$, 
it is possible to get a bilayer plasmon feature though the 
superconducting delta function is absent.  

%\vspace{0.25cm} 
\centerline{\epsfxsize=3.2truein \epsfbox{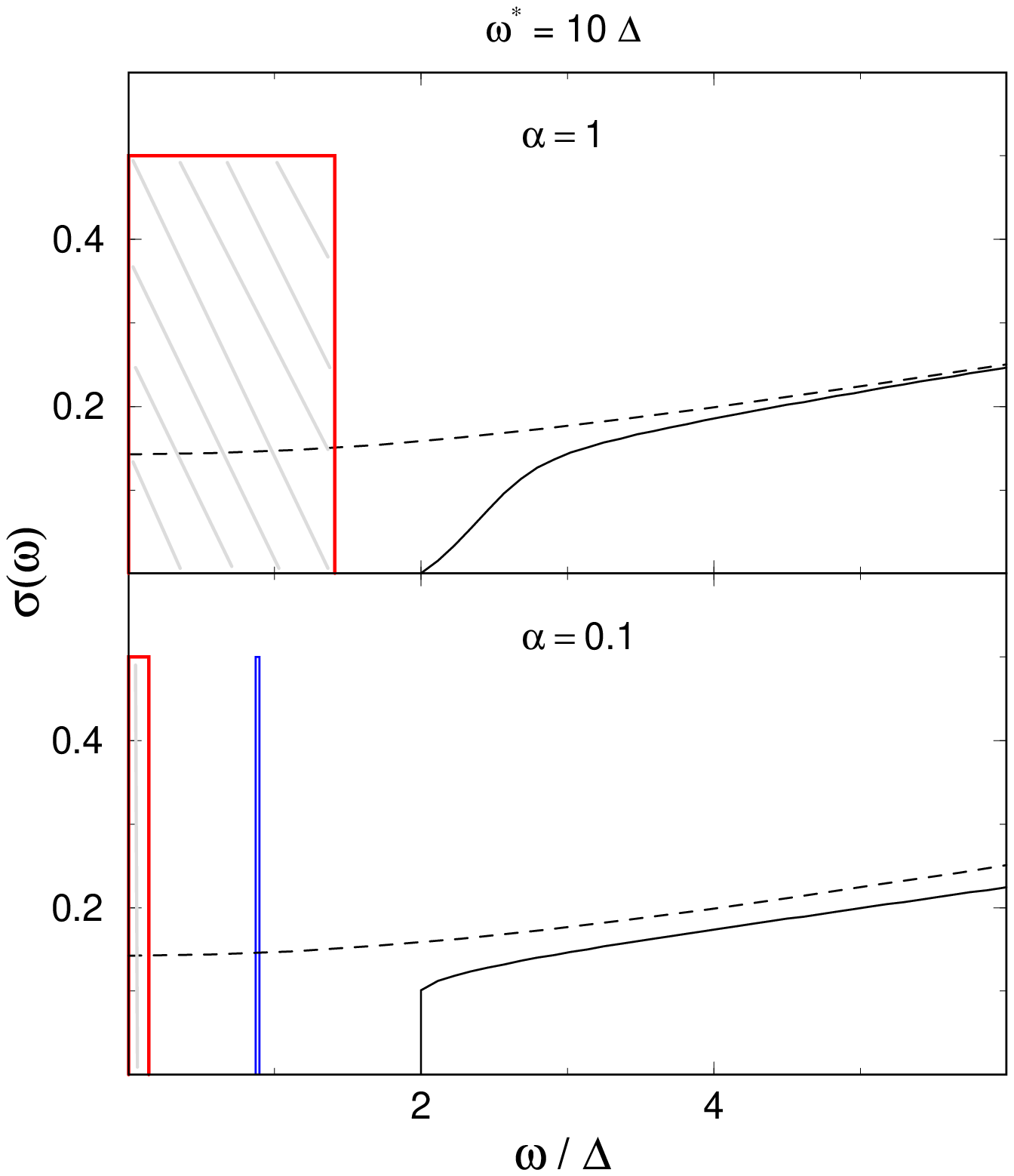}}
{\footnotesize \textbf{Fig. 6} Real part of conductivity for 
bilayer $(b=0.075)$ system for $\omega^* = 10 \Delta$. 
Top panel corresponds to fully phase coherent ($\alpha_\nu = 1$) and  
bottom panel to $\alpha_{\nu}=0.1$  superconducting state. The dashed 
line in both  corresponds to the normal state conductivity. The area 
of the shaded rectangle at $\omega=0$ in each panel equals 
$\pi \rho_s /s$ while the area of the rectangle at 
$\omega_{bilayer}$ in the second panel  equals $\pi \rho_{bilayer}$. }
\vspace{0.25cm}

Fig 7 plot the bilayer plasmon
frequency $\omega_{bilayer}$, the spectral weight in the bilayer plasmon 
$\rho_{bilayer}$ (defined by Eq \ref{rhob}) and the spectral weight in 
the superconducting delta function(at $\omega = 0$) $\rho_s$ 
as a function of $\alpha_1= \alpha_2$ for $b=0.1$ and $\omega^{*}=0.4 \Delta$. The two spectral weights vary linearly in the fluctuation parameter for the 
given value of $\omega^{*}$ and hence their ratio is independent of the $\alpha_{\nu}$ value.

\vspace{0.25cm} 
\centerline{\epsfxsize=3.2truein \epsfbox{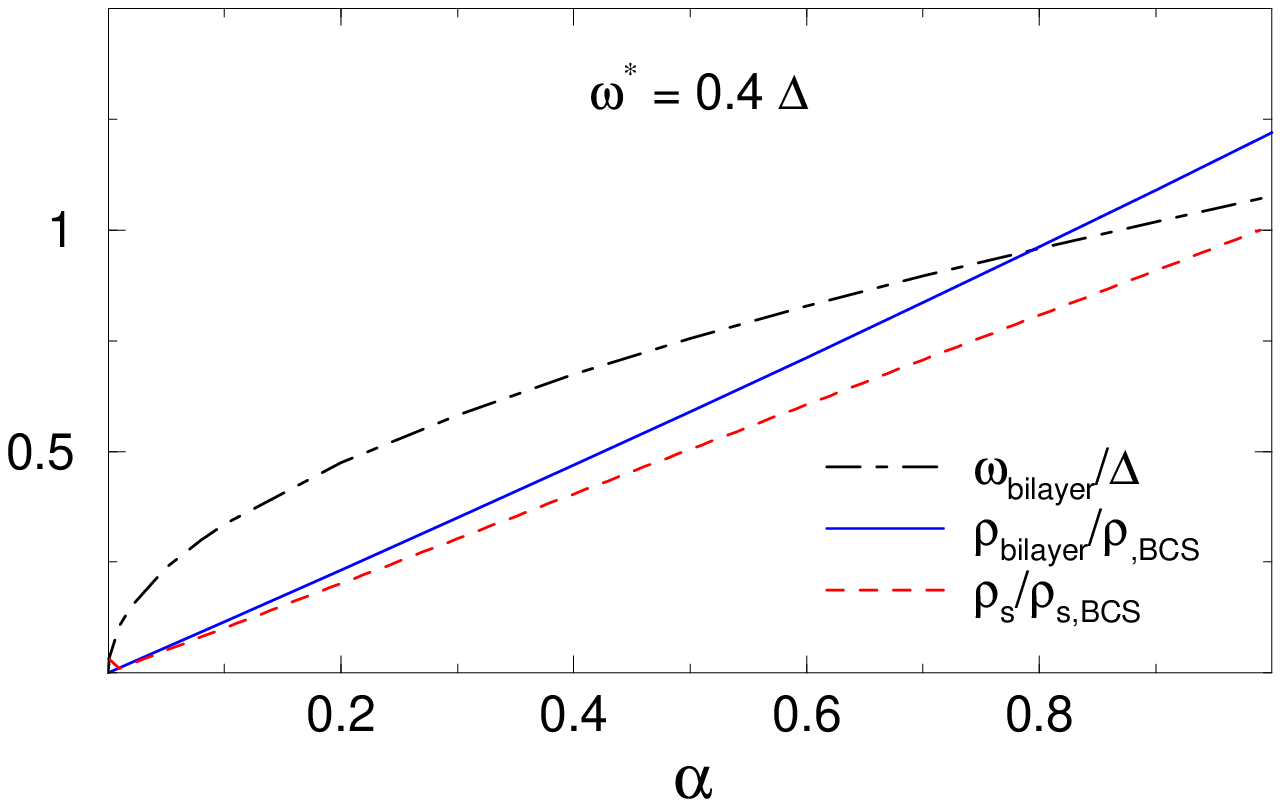}}
{\footnotesize \textbf{Fig. 7} The bilayer plasmon
frequency $\omega_{bilayer}$, the spectral weight in the bilayer 
plasmon $\rho_{bilayer}$ (defined by Eq \ref{rhob}) and the spectral 
weight in the superconducting delta function (at $\omega = 0$) 
$\rho_s$ as a function of $\alpha_1= \alpha_2$ for $b=0.075$ and $\omega^{*}=0.4 \Delta$.  }
\vspace{0.25cm}

\section{Superfluid Stiffness and Sum Rules}

At very low frequencies in the superconducting state one has 
\begin{equation}
\sigma _{bilayer}\rightarrow \frac{i \rho _{s,bilayer}}{\omega }.
\end{equation}
Inspection of Eq \ref{sigmabilayer} shows that 
\begin{equation}
\rho _{s,bilayer}=\frac{\rho _{s,1}\rho _{s,2}}{\rho _{s,1}\widetilde{d}%
_{2}+\rho _{s,2}\widetilde{d}_{1}}=\pi \sigma _{0,1}\Delta \frac{b\alpha
_{1}\alpha _{2}}{\alpha _{1}\widetilde{d}_{2}+b\alpha _{2}\widetilde{d}_{1}},
\label{rhosbilayer}
\end{equation}
where the first equation applies to all bilayer systems and the second
follows from the specific assumptions we have made. One question of current
experimental interest is the relation between the superfluid stiffness and
changes in conductivity as the temperature is reduced (a) below the
`pseudogap' scale at which the gap opens and (b) below $T_{c}$ at which
phase coherence is established. In  Refs \cite{Ioffe99,Ioffe00} these
relations were established for the `single-layer' case.
The numerical results presented in the previous section show that differences 
occur in the bilayer case. To analyze this issue more precisely, we note 
that the dirty-limit model analyzed in this paper should be viewed as 
arising from a model with a very large but finite
scattering rate $\Gamma $ in the limit $\left( \Omega
,\Omega _{bilayer},\Delta \right) << \Gamma $. The standard sum rule 
derivations are based on analysis of the $\Omega /\Gamma \rightarrow \infty $ limit. 
However, one may obtain sum rules for the superconductivity and pseudogap 
induced changes in $\sigma$ without considering this limit. We define the 
change in the spectral weight as $\Delta$ is increased from $\Delta =0$ by 
\begin{equation}  
\delta K(\Omega, \Delta, \alpha)=\int_{0}^{\Omega} 
\frac{2d\omega }{\pi }\left[\sigma(\Delta, \alpha ) -\sigma(\Delta = 0) \right].
\label{K}
\end{equation}
This quantity remains finite in the limit $\Omega \rightarrow \infty$, $\Omega /\Gamma << 1$. 
The values we obtain with our $\Gamma  \rightarrow \infty$ limit are 
accurate up to terms of relative order  $\left(\Omega _{bilayer},\Delta \right) /\Gamma $. 

It is also useful to consider the change 
in the spectral weight excluding the superfluid response : we define 
\bea
\delta K_{+}(\Omega, \Delta, \alpha) &=& \int_{0^{+}}^{\Omega }\frac{2d\omega }{\pi }
[\sigma(\Delta, \alpha ) -\sigma(\Delta = 0)] \nonumber\\
&=& \delta K(\Omega, \Delta, \alpha) - \rho_s (\Delta, \alpha) 
\label{K+}
\eea
and the ratio of change in $\delta K_{+}$ with $\alpha$ to the superfluid stiffness given by
\bea
R(\Omega) = \frac{\delta K_{+}(\Omega , \Delta, \alpha) - \delta K_{+}(\Omega, \Delta,0)}
{\rho_s (\Delta, \alpha)}.
\label{R}
\eea

Now Refs \cite
{Ioffe99,Ioffe00} showed that for a `single-layer'(s-l) system, the change in the 
spectral weight as defined by Eq \ref{K} is 
\bea
\delta K_{s-l}(\Omega = \infty, \Delta, \alpha) 
 = -\frac{(1-\alpha)\rho_s^{BCS}(\Delta )}{2},
\eea
where $\rho _{s}^{BCS}(\Delta )$ is defined by Eq \ref{rhoBCS}.
Thus the change with $\alpha$ in total spectral weight at fixed $\Delta$ is 
\bea
&&\delta K_{s-l}(\Omega = \infty, \Delta, \alpha) - \delta K_{s-l}(\Omega
= \infty, \Delta,0) \nonumber \\
&& =\frac{\alpha \rho _s^{BCS}(\Delta )}{2}
\eea
and the change with $\alpha$ in the $\Omega >0$ spectral weight as defined by Eq \ref{K+} is 
\bea
&&\delta K_{+,s-l}(\Omega = \infty, \Delta, \alpha) - \delta K_{+,s-l}
(\Omega = \infty, \Delta,0) \nonumber\\
&&=-\frac{\alpha \rho _{s}^{BCS}(\Delta)}{2} = -\frac{\rho_{s}^{s-l}(\Delta, \alpha)}{2}
\eea
Use of Eq. 21 gives the value of $R_{s-l}(\Omega = \infty) = -1/2$.
In other words, if the superconducting gap appears without phase coherence,
the oscillator strength decreases by an amount related to the `BCS'
superfluid stiffness, essentially because the conductivity in the region
less than the gap is suppressed and no additional oscillator 
strength appears in the superfluid response. If phase coherence is now turned on, the
total oscillator strength and the superfluid stiffness increase, while the
spectral weight in the $\Omega >0$ conductivity decreases.  
Comparison of Eq 14 and Eq 23 shows that in the `single-layer' case the ratio between these changes
is 2:1.

Applying these arguments to the bilayer case shows that
\begin{equation}
\delta K_{bilayer}(\Omega = \infty, \Delta ,\alpha_{\nu} )= -\frac{1}{2}
\sum_{\nu }(1-\alpha_{\nu })\pi\sigma _{0,\nu }\widetilde{d}_{\nu}\Delta .
\label{K_bilayer}
\end{equation}

The simple factor-of-two relation between the 
phase-coherence-induced change in $\delta K_{+}$ 
and the superfluid stiffness does not
occur in the bilayer system  
essentially because when the phase coherence parameter is varied, the 
strengths of both the bilayer plasmon feature and the superfluid stiffness vary. 
To see what the relation is, we plot in Fig 8 the ratio $R_{bilayer}(\Omega = \infty) $  
as a function of bilayer anisotropy b ($R_{bilayer}(\Omega = \infty)$ being 
independent of $\alpha_1= \alpha_2 = \alpha$). We see that the ratio increases 
monotonically as $b$ is decreased from the single-layer value $b = 1$, 
and changes sign at a $d_1/d_2$ dependent value of $b \sim 0.2$. 
Thus unlike in the single-layer case, where $\rho_s$ increased by twice 
the decrease in $\delta K_{+}(\Omega = \infty)$, in the bilayer case the 
increase is generically greater than $2\delta K_{+}(\Omega = \infty)$ 
and indeed for extreme anisotropy both $\delta K_{+}(\Omega = \infty)$ and $\rho_s$ 
increase as $\alpha$ is increased from zero. 

\vspace{0.25cm} 
\centerline{\epsfxsize=3.2truein \epsfbox{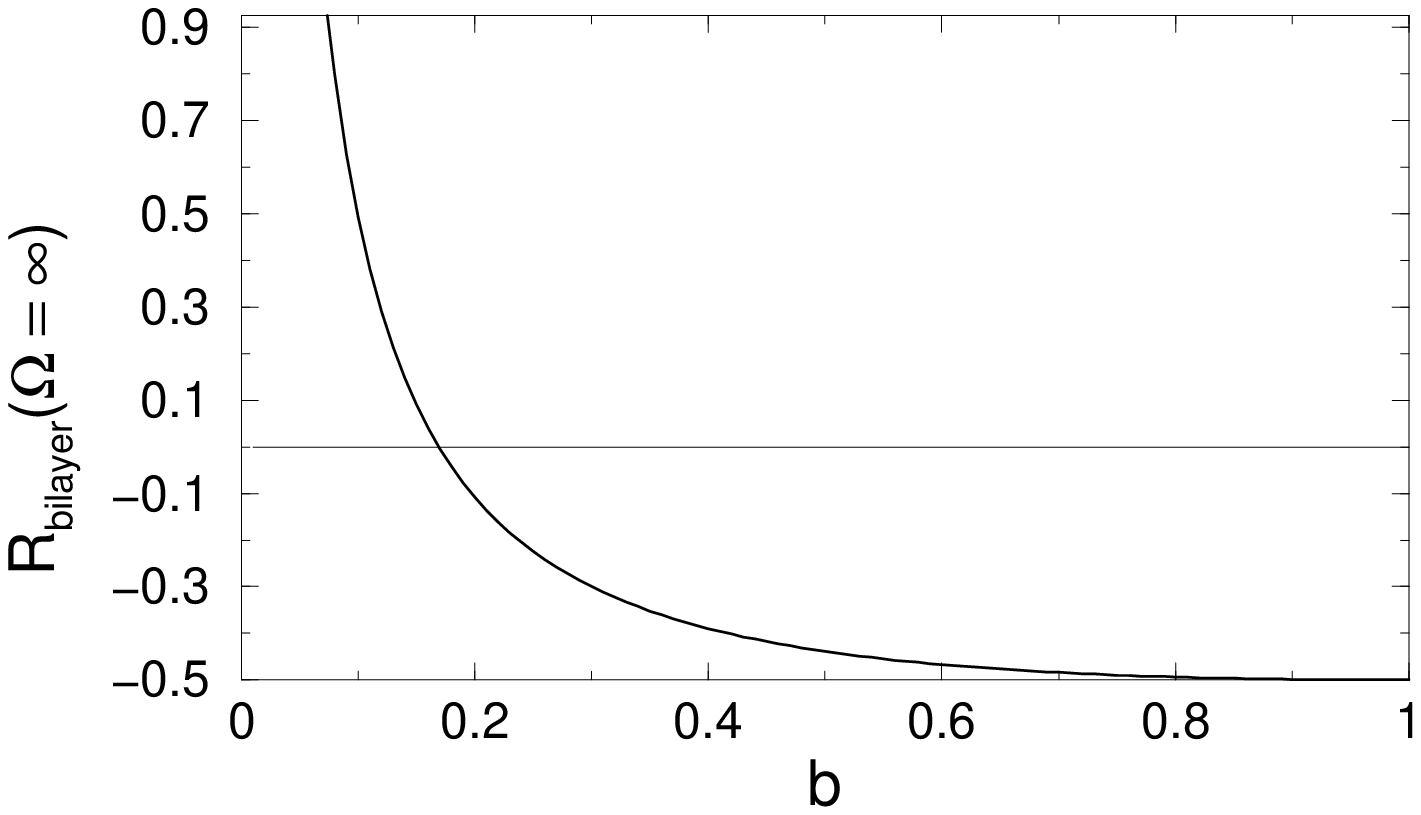}}
{\footnotesize \textbf{Fig. 8} Ratio of the change in $\delta K_{+}$ 
on onset of phase coherence to the superfluid stiffness defined by 
Eq \ref{R}, $R_{bilayer}(\Omega = \infty)$ as a function of the 
anisotropy parameter $b$. Note that the ratio is independent of 
the value of $\alpha$ and that the $b=1$ value corresponds to the single-layer case.   }
\vspace{0.25cm}

Fig 9 plots $R_{bilayer}$ as  a 
function of the cut-off $\Omega$. 
The $\Omega = \infty$ value is indicated by an arrow. We have verified
that the calculated quantity does indeed converge to the 
correct $\Omega \rightarrow \infty$ value, but as can be seen, the convergence
is very slow.

\vspace{0.25cm} 
\centerline{\epsfxsize=3.2truein \epsfbox{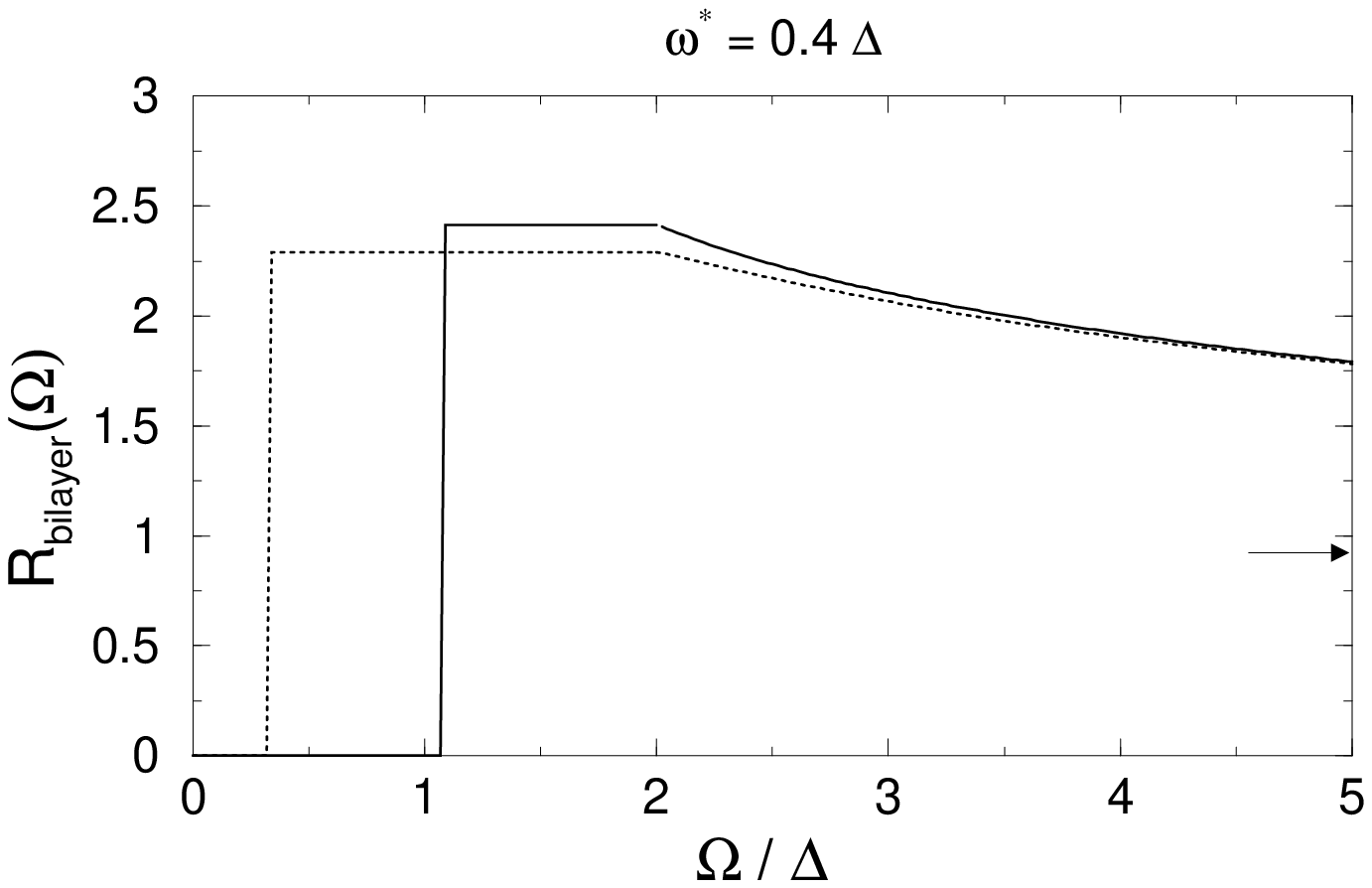}}
{\footnotesize \textbf{Fig. 9} Ratio of the change in $\delta K_{+}$ 
on onset of phase coherence to the superfluid stiffness 
defined by Eq \ref{R}, $R_{bilayer}$ plotted as a function of 
cut-off $\Omega$ for a bilayer $(b=0.075)$ system with $\omega^* =0.4 \Delta$.  
Solid line: fully phase coherent ($\alpha_\nu = 1$) superconducting state;
dotted line: $\alpha_{\nu}=0.1$  superconducting state.
Arrow: $R_{bilayer}(\Omega = \infty) $ .}
\vspace{0.25cm}

Fig 10 shows the changes in spectral weight for the case of a 
large difference in interplane hopping $(b=0.075)$ and for $\omega^{*} = 0.4 \Delta$ 
as a function of the cut-off frequency $\Omega$.  For $\alpha = 1$ (top panel), we 
expect conservation of the total spectral weight while for $\alpha = 0.1$ (bottom panel) 
we expect the  $\Omega \rightarrow \infty$ value to be non-zero as given 
by Eq \ref{K_bilayer}. A remarkably slow convergence of the change in the 
spectral weight to its $\Omega \rightarrow \infty$ value is evident. We verify 
the $\Omega \rightarrow \infty$ values numerically by plotting in Fig 11, 
$\delta K_{bilayer}$ as a function of inverse cut-off frequency. We have 
verified that the $1/\Omega \rightarrow 0 $ limit matches the value given 
by Eq \ref{K_bilayer}. Considerable caution must be exercised in the experimental 
investigations of changes in the spectral weight because small differences 
persisting over wide frequency ranges may lead to appreciable contributions to sum rules.

\vspace{0.25cm} 
\centerline{\epsfxsize=3.2truein \epsfbox{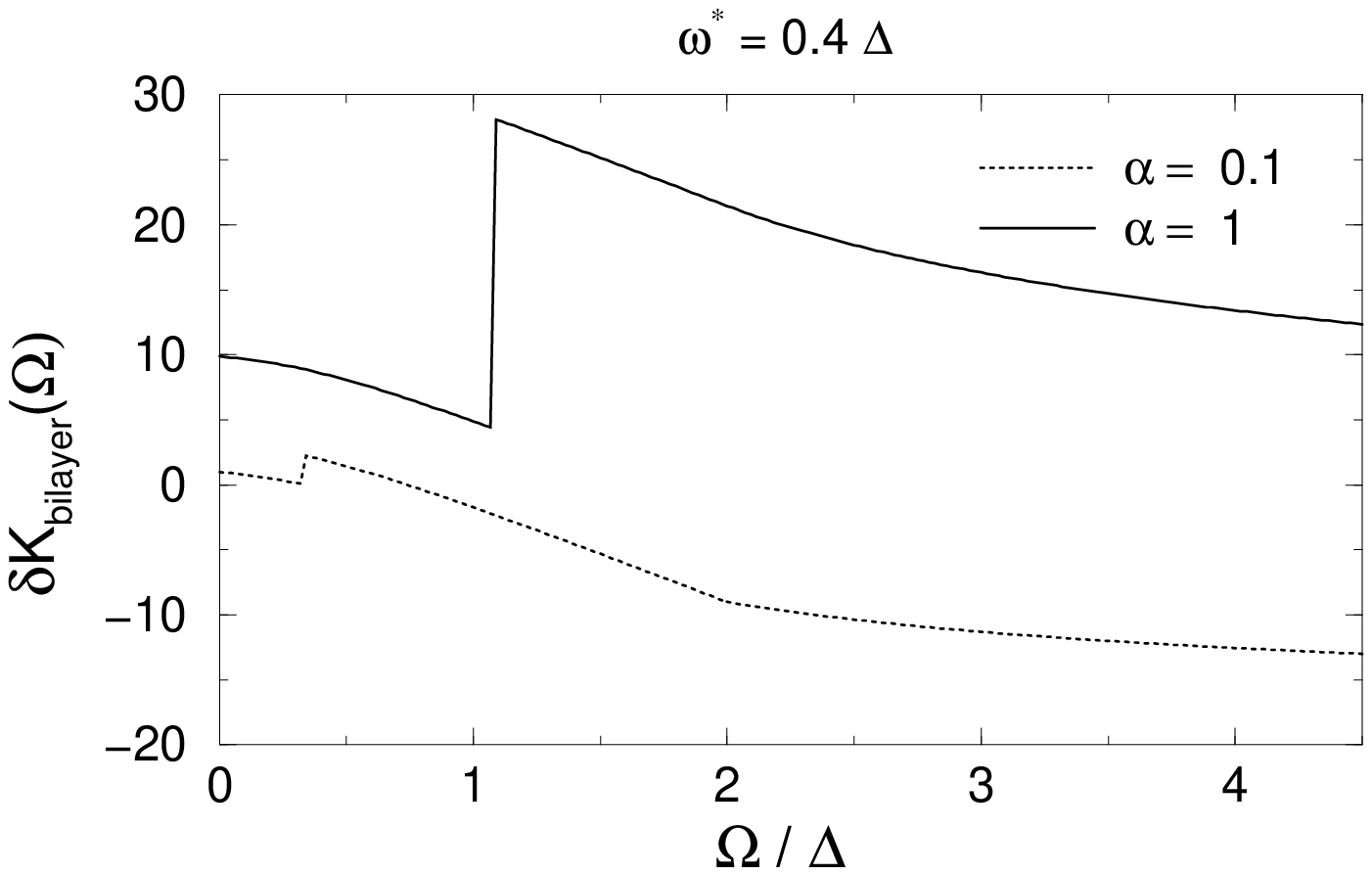}}
{\footnotesize \textbf{Fig. 10} $\delta K_{bilayer}$ 
(defined by Eq \ref{K}) as a function of cut-off frequency 
$\Omega$ for bilayer $(b=0.075)$ system for $\omega^* =0.4 \Delta$. 
Solid line corresponds to fully phase coherent ($\alpha_\nu = 1$) and  
dotted line to $\alpha_{\nu}=0.1$  superconducting state. }
\vspace{0.25cm}

\vspace{0.25cm} 
\centerline{\epsfxsize=3.2truein \epsfbox{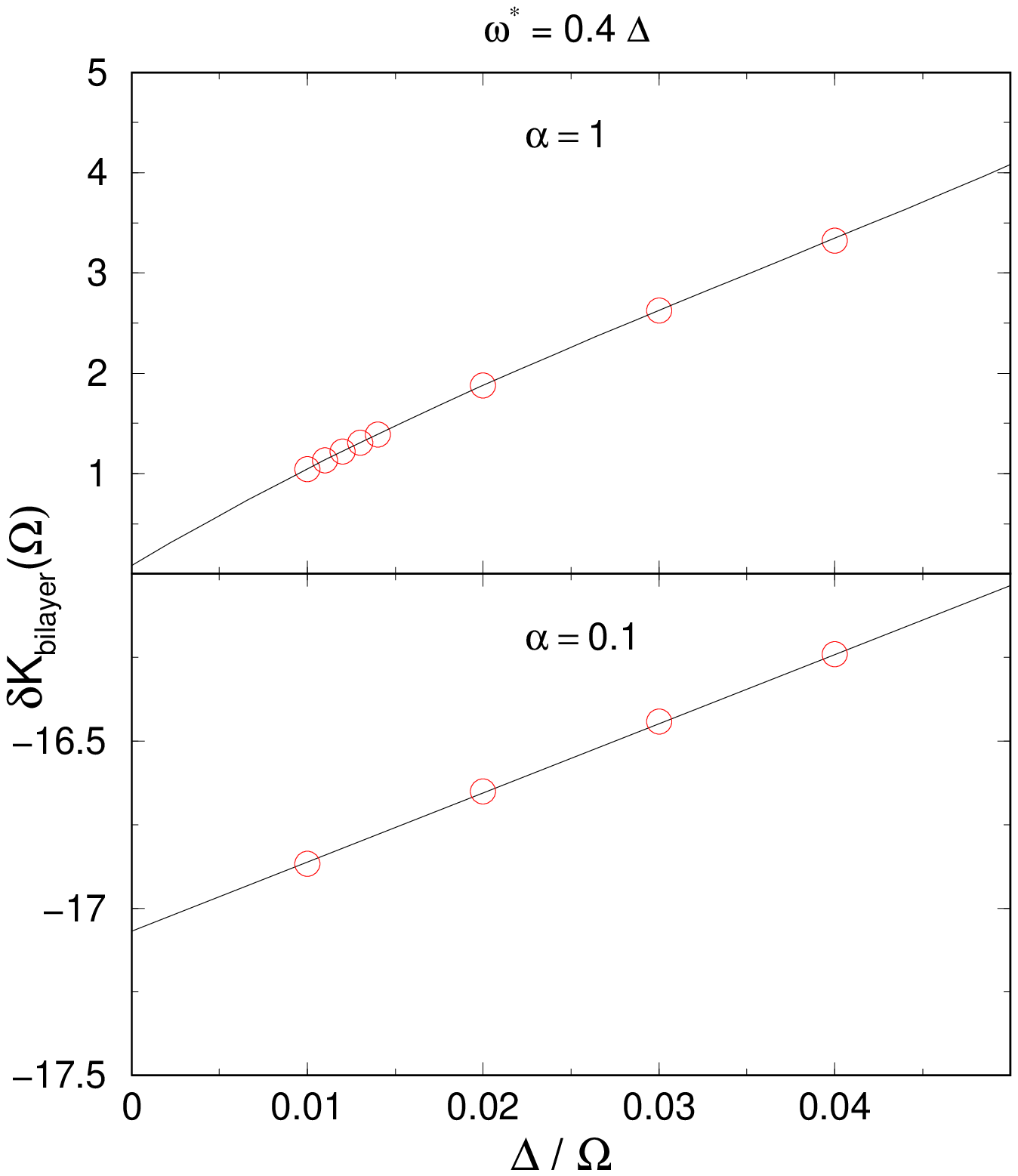}}
{\footnotesize \textbf{Fig. 11} $\delta K_{bilayer}$ as a 
function of inverse cut-off frequency $\Omega$ demonstrating
convergence to correct sum-rule value. The $1/\Omega = 0$ 
value correctly gives the $\Omega \rightarrow \infty$ value as given 
by Eq \ref{K_bilayer}. Top panel corresponds to fully phase coherent 
($\alpha_\nu = 1$) and  bottom panel to $\alpha_{\nu}=0.1$  superconducting state.   }

\vspace{0.25cm}

\section{Inclusion of phonons}

Due to the proximity of the bilayer feature to the optical phonons 
in high-$T_c$ materials like $Bi_{2}Sr_{2}Ca_{2}Cu_{2}O_{8}$
and $YBa_{2}Cu_{3}O_{7-\delta }$, it is of interest to study the 
interplay of phonons and the bilayer plasmon. In the first part of this paper
we had taken $\varepsilon_1 = \varepsilon_2= \varepsilon_{\infty} = const$. 
To  incorporate phonon modes we include in our analysis frequency dependent dielectric
functions for each layer:
$\varepsilon_1(\omega)$ and $\varepsilon_2(\omega)$. We
obtain for interplane bilayer conductivity,

\bea
\sigma_{bilayer}^{phonon}(\omega ) = &&\frac{\sigma _{1}\sigma _{2}-\frac{i\omega}
{4 \pi} (\sigma
_{1}(\varepsilon_2-\widetilde{d}_{2}) +\sigma _{2}(\varepsilon_1- \widetilde{d}_{1}))}
{\sigma _{1}\widetilde{d}_{2}+\sigma _{2}\widetilde{d}
_{1}-\frac {i\omega}{4 \pi}(\varepsilon_1\widetilde{d}_{2}  + \varepsilon_2 \widetilde{d}_{1})} 
\nonumber \\
&&- \frac{\frac{\omega^2}{16 \pi^2}(\varepsilon_1 \varepsilon_2 - 
(\varepsilon_1\widetilde{d}_{2} + \varepsilon_2 \widetilde{d}_{1}) )}{\sigma _{1}\widetilde{d}_{2}+\sigma _{2}
\widetilde{d}_{1}-\frac {i\omega}{4 \pi}(\varepsilon_1\widetilde{d}_{2}  + 
\varepsilon_2 \widetilde{d}_{1})}.  
\label{sigmabilayerphonon}
\eea
This more general formula reduces to Eq \ref{sigmabilayer} 
with a value of $C = 4 \pi / \varepsilon$ on choosing  
$\varepsilon_1 = \varepsilon_2= constant$ (inclusion of the 
compressibility term $\chi$ leads to more complicated formulae).

To study the qualitative effect of including phonons we consider the 
simplest possible case of $\epsilon_2=1$ and
\bea
\varepsilon_1(\omega) = \varepsilon_{\infty} + \frac{(\varepsilon_{0} - 
\varepsilon_{\infty}) \omega_p^2}{\omega_p^2 - \omega^2 - i \omega \gamma}.
\eea   
where $\omega_p$ is the frequency of the phonon mode and $\gamma$ the broadening. 

Fig. 12 displays the effects of adding, to the situation 
(full phase coherence ($\alpha=1$) and bilayer plasmon inside the gap) shown in 
the top panel  of Fig 4,
a phonon with a frequency greater than
the bilayer plasmon frequency. For orientation, the top panel shows the bilayer plasmon part of
the electronic absorption in the absence of phonons (calculated from Eq.\ref{sigmabilayerphonon} with $\varepsilon_1 = \varepsilon_2= 1$)  , and the phonon absorption in the absence
of electrons (calculated from Eq. \ref{sigmabilayerphonon} with $\sigma_1=\sigma_2=0$). The bilayer
plasmon was represented in Fig 4 as a rectangle and is shown here with a Lorentzian broadening. The phonon feature
is rendered optically active and shifted up from the phonon frequency $\omega_p$ by bilayer effects.
The lower panel of Fig 12 shows the full conductivity. The electronic continuum contribution
at $\omega>2\Delta$ is present but very difficult to perceive on the scale of this plot. It 
is evident that the coupling between the modes leads as usual to level repulsion, and that
further, almost all of the oscillator strength goes into the upper mode. The lower mode
(shown in expanded view in the inset) is almost invisible. 

\vspace{0.25cm} 
\centerline{\epsfxsize=3.2truein \epsfbox{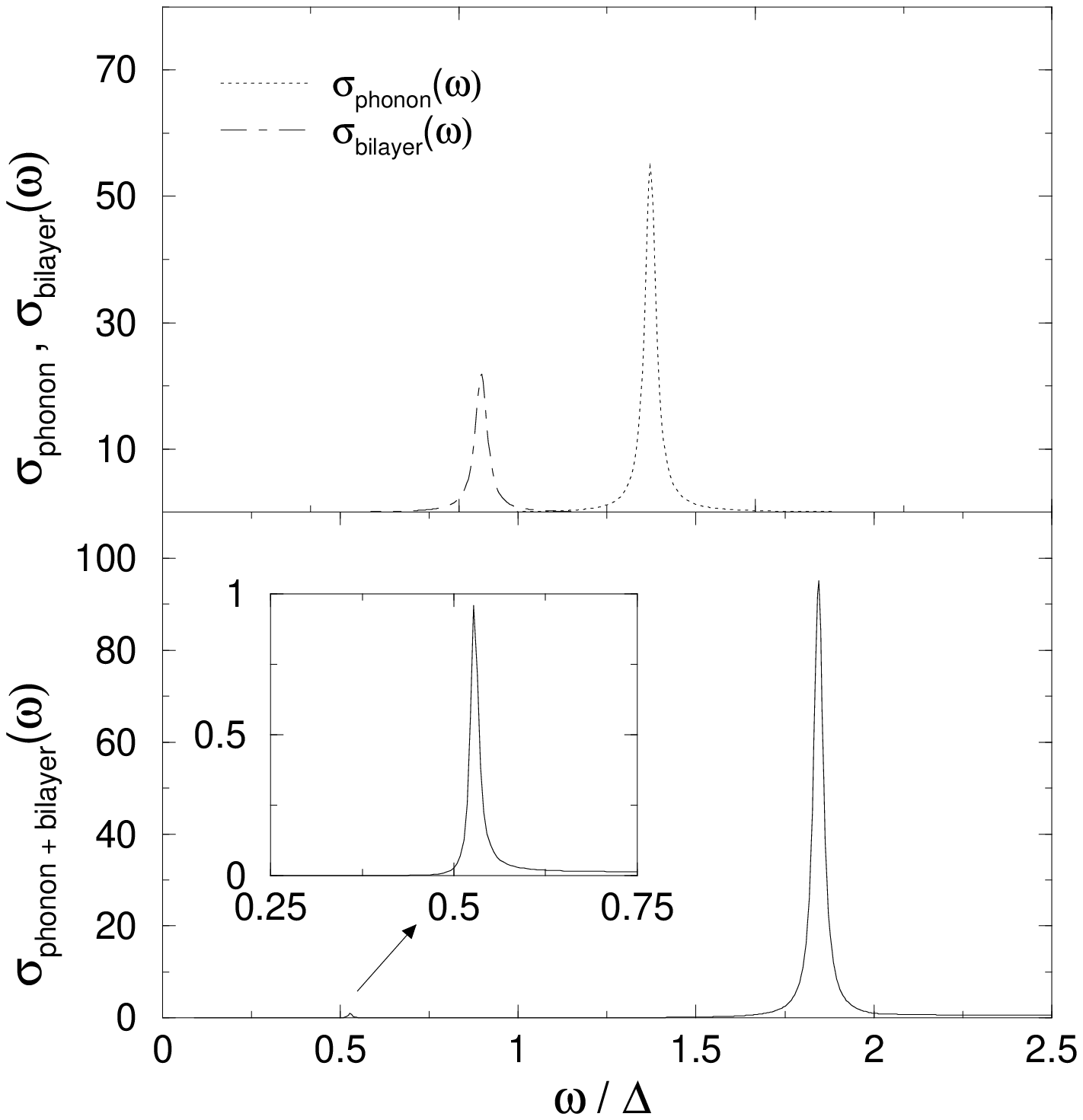}}
{\footnotesize \textbf{Fig. 12} Top panel plots the conductivity 
$\sigma_{phonon}(\omega )$ with  $\omega_p = 0.91 \Delta$ obtained from 
Eq \ref{sigmabilayerphonon} by putting the electron conductivities 
$\sigma _{1,2}=0$ (dotted line) and the broadened bilayer conductivity  
$\sigma_{bilayer}(\omega )$ with $\omega^* = 0.4 \Delta$ and $b = 0.075$ (dot-dashed line). The bottom panel 
plots $\sigma_{bilayer}^{phonon}(\omega )$ (Eq \ref{sigmabilayerphonon}). 
The inset shows the lower peak not visible on the scale of the plot.  }
\vspace{0.25cm}

\vspace{0.25cm} 
\centerline{\epsfxsize=3.2truein \epsfbox{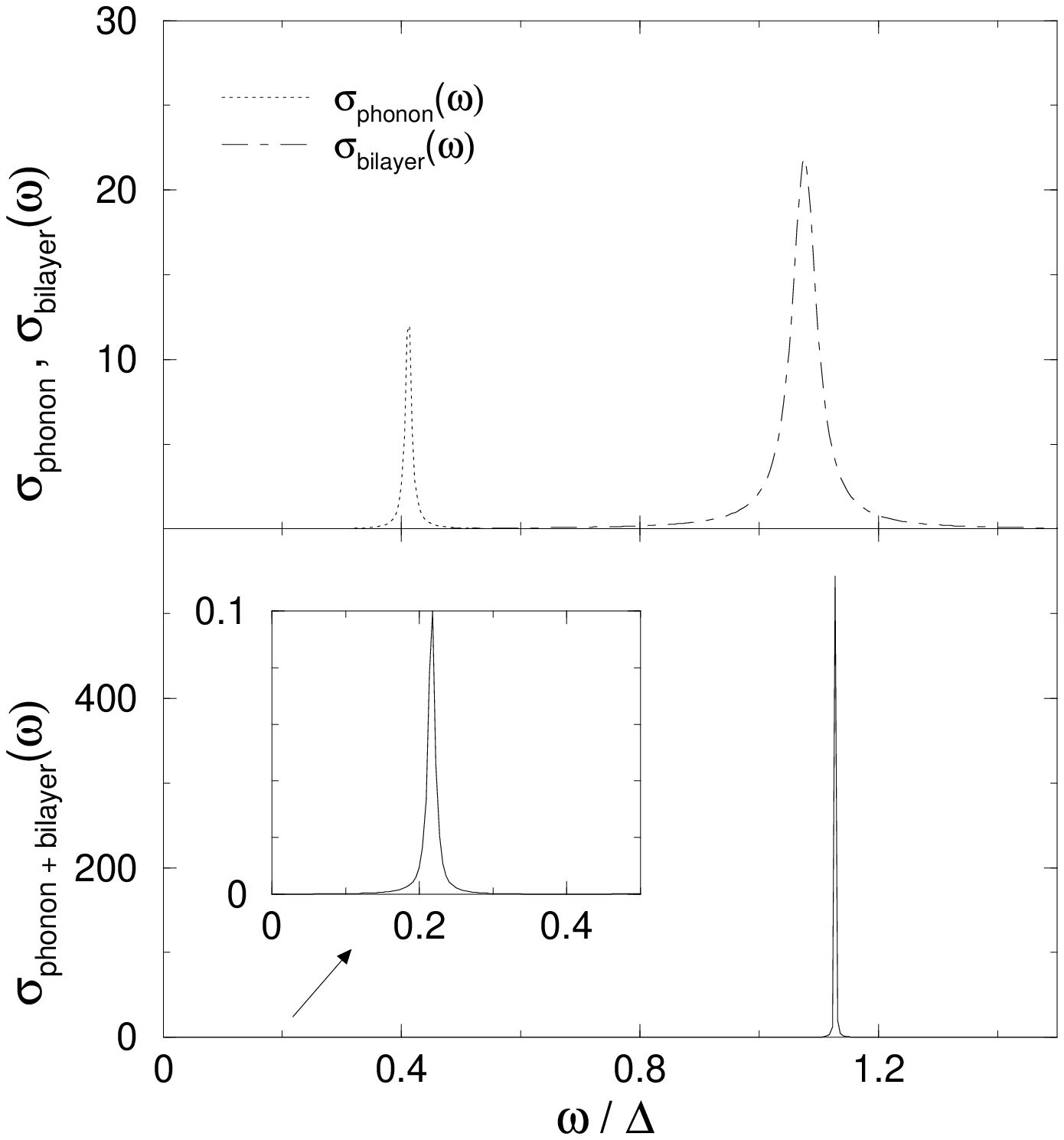}}
{\footnotesize \textbf{Fig. 13} Top panel plots the conductivity 
$\sigma_{phonon}(\omega )$ with  $\omega_p \sim 0.23 \Delta$ obtained 
from Eq \ref{sigmabilayerphonon} by putting the electron conductivities 
$\sigma _{1,2}=0$ (dotted line) and the broadened bilayer conductivity  
$\sigma_{bilayer}(\omega )$ with $\omega^* = 0.4 \Delta$ and $b = 0.075$ 
(dot-dashed line). The bottom panel plots 
$\sigma_{bilayer}^{phonon}(\omega )$ (Eq \ref{sigmabilayerphonon}). The inset 
shows the lower peak not visible on the scale of the plot.  }
%\vspace{0.25cm}

Fig. 13 shows that roughly the same situation is obtained if the phonon starts out at a lower frequency  
than the bilayer plasmon. In this case the upper mode shifts by rather less, but the qualitative
features are the same.

We now briefly outline the effects of increasing phase fluctuations
(decreasing $\alpha$ from unity) i.e increasing temperature. As can be seen from Figs 4-6, increasing
phase fluctuations decreases the frequency and oscillator strength of the bilayer plasmon mode. Thus if the `bare' phonon frequency is greater than
the bilayer plasmon frequency (as in Fig 12), then relatively minor changes occur in the absorption spectrum as $\alpha$ is decreased.  Essentially, the almost invisible lower absorption moves to the left and becomes a bit sharper which in turn results in the slight decrease in frequency and intensity of the upper absorption.

\vspace{0.25cm} 
\centerline{\epsfxsize=3.2truein \epsfbox{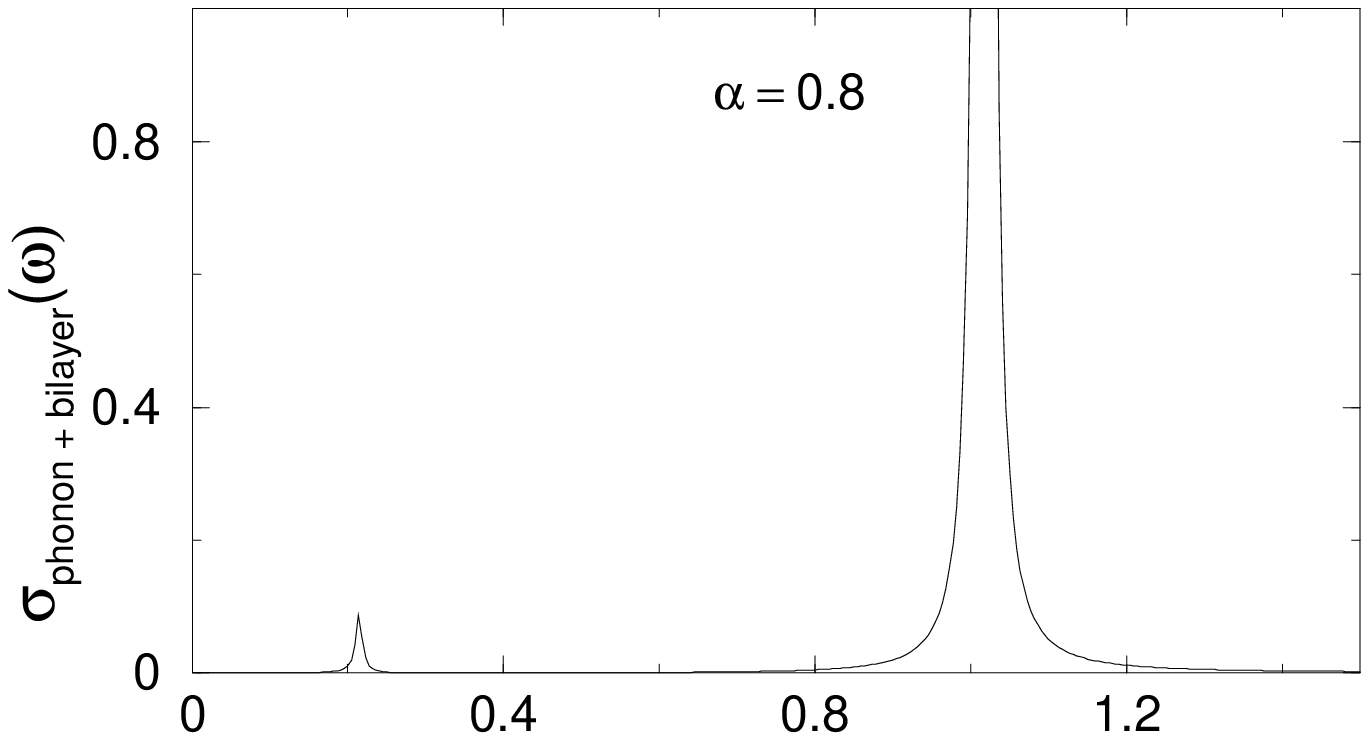}}
\vspace{0.05cm} 
\centerline{\epsfxsize=3.2truein \epsfbox{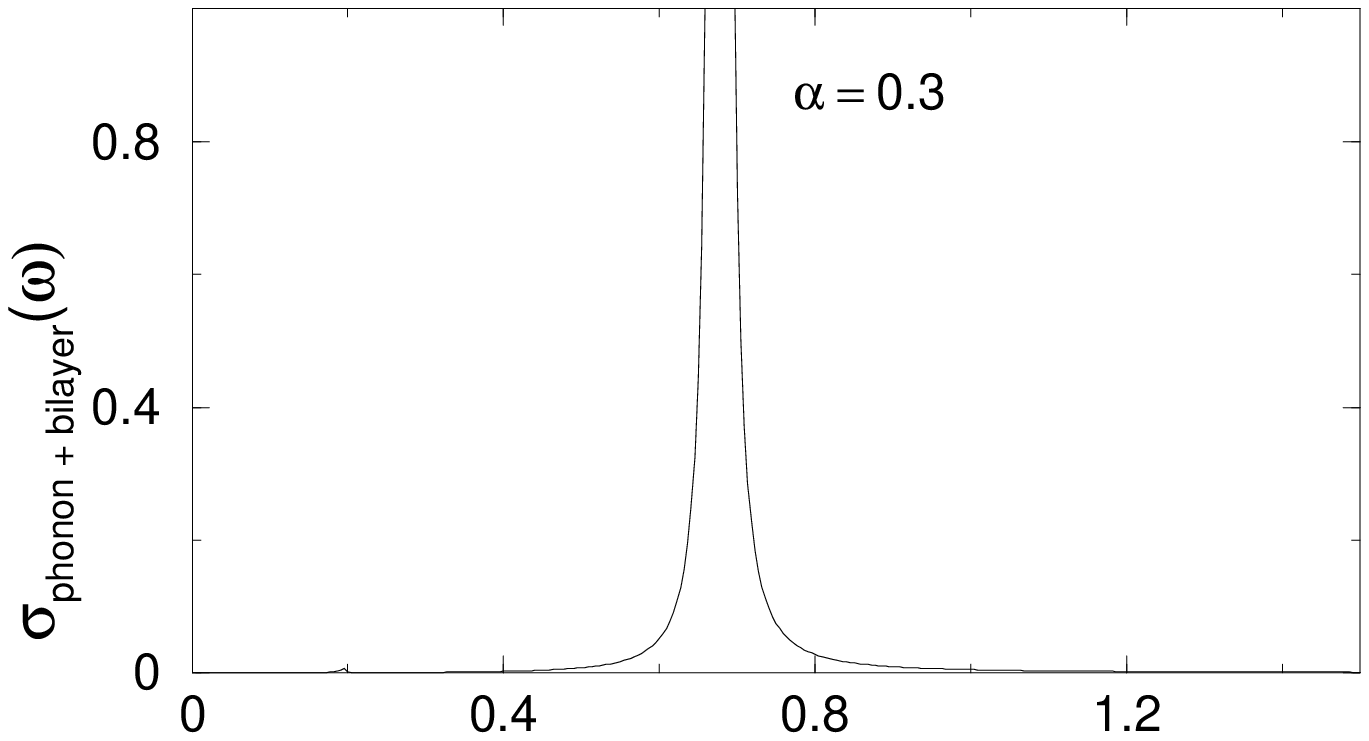}}
\vspace{0.05cm} 
\centerline{\epsfxsize=3.2truein \epsfbox{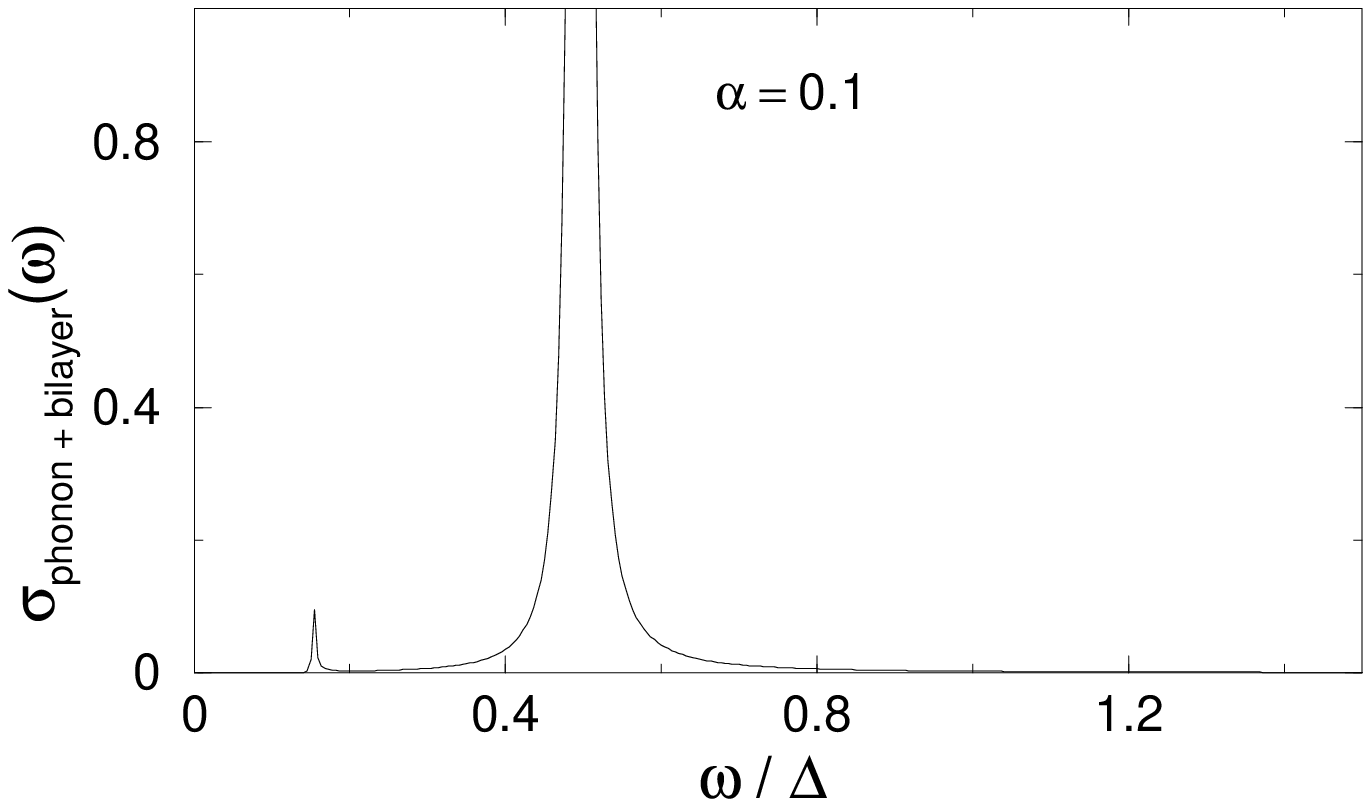}}
{\footnotesize \textbf{Fig. 14} The figure plots 
$\sigma_{bilayer}^{phonon}(\omega )$ (Eq \ref{sigmabilayerphonon}) for $\alpha = 0.8$ (top panel), $\alpha = 0.3$ (panel two) and $\alpha = 0.1$ (bottom panel)  for the same set of parameters as in Fig 13 ($\omega_p \sim 0.23 \Delta$ $\omega^* = 0.4 \Delta$). }
\vspace{0.25cm}

On the other hand, if the $\alpha=1$
bilayer plasmon is at a higher frequency than the phonon, more drastic changes will occur as shown in Fig 14. Decreasing $\alpha$ slightly results  in a decrease in frequency and intensity of the stronger upper absorption as shown in the top panel of Fig 14 for $\alpha = 0.8$. As $\alpha$ is further decreased $\alpha$ the $\sigma_{bilayer}$ peak in Fig 13 goes on moving to the left and eventually overlaps with the $\sigma_{phonon}$ peak at which point the lower absorption is almost invisible and the conductivity is as shown in the second panel of Fig 14 for $\alpha = 0.3$. On further decreasing $\alpha$ the  $\sigma_{bilayer}$ peak crosses over to the left of  $\sigma_{phonon}$ and the the conductivity is as given by the bottom panel of Fig 14 for  $\alpha = 0.1$. Also note that the decrease in intensity of the upper peak is much stronger with the decrease in $\alpha$ as compared to the case when the `bare' phonon frequency is greater than
the bilayer plasmon frequency.  From experimental point of view, the behavior of the observed peaks as a function of temperature might allow us to distinguish between the two situations where the `bare' phonon frequency is greater than or smaller than the bilayer plasmon frequency. 

Finally, we note  that the spectral weight in the
superfluid stiffness is unaffected by the phonons.

\section{Conclusions and Applications to experiment}

We have extended the theory of refs \cite{Ioffe99,Ioffe00} to the bilayer situation of 
relevance to many experimentally studied high-$T_c$ materials. The crucial physics is 
local-field or ``blockade'' effects: the difference in hopping amplitudes 
characteristic of a bilayer structure leads to charge imbalances inside 
the unit cell; the electrochemical potentials due to these charge imbalances act 
to suppress the low frequency response to a uniform electric field, and lead to 
bilayer plasmon features in the absorption. For the physics of high-$T$ the crucial
question is the observability of the effects of thermal and quantal
fluctuations of the phase of the superconducting order parameter,
here parameterized by a 'Debye-Waller parameter $\alpha$

We find a low frequency suppression of the conductivity in the normal (neither 
superconducting or pseudogap) state discussed in section III and shown in Fig 2. 
However this distinct signature is not apparent  in the experimental plots of 
the normal state conductivity. There are two possibilities: either the scale 
is very high or the effect is masked by phonons. 

A second, generally valid qualitative result is that the simple factor of two 
relation between $\rho_s$ and the change with onset of phase coherence in the 
$\omega > 0$ oscillator strength which was found for single-layer systems, no 
longer applies for bilayer systems (cf Eq \ref{K_bilayer} and Fig 8); the 
change in $\rho_s$ is generically
 greater. This qualitative behavior has been observed.
 
A third important result is that the convergence of sum-rule integrals with 
frequency can be very slow; so caution should be exercised in applying sum-rule 
arguments to data.

Further we showed that the coupling of phonons, phase fluctuations and the bilayer 
plasmon leads to complicated effects on the spectrum, which depend sensitively on 
parameters suggesting that an unambiguous extraction of the bilayer 
plasmon frequency and spectral weight may be difficult. This is unfortunate, as the 
these quantities in principle carry information about the physically crucial phase
fluctuation properties encoded in the Debye-Waller parameter $\alpha$. We suggest however how the change is conductivity with increasing phase fluctuations (or increasing temperature) is expected to be different based on whether the phonon frequency is above or below the bilayer plasmon frequency.

{\it Acknowledgement: }This work was supported in part by NSF-DMR-00081075 and stemmed from a crucial remark of L. B. Ioffe, whose advice and insight we gratefully acknowledge.
AJM thanks L. B. Ioffe, S. Das Sarma and D. van der Marel (who pointed out an error in a previous version)  for helpful conversations and
the Institute for Theoretical Physics and Brookhaven National Laboratories
for hospitality and support.

\end{multicols}
\end{document}